\documentclass[aps,prb,showpacs]{revtex4}
\usepackage{epsfig}
\begin{document}
\title{Coulomb drag in high Landau levels}
\author{I.V.\ Gornyi$^{1,\dagger}$, A.D.\ Mirlin$^{1,2,\ddagger}$, and
F.\  von Oppen$^{3,4}$ }
\affiliation{$^1$Institut f\"ur Nanotechnologie, Forschungszentrum
  Karlsruhe, 76021 Karlsruhe, Germany\\
$^2$Institut f\"ur Theorie der Kondensierten Materie,
Universit\"at Karlsruhe, 76128 Karlsruhe, Germany\\
$^3$Department of Condensed
Matter Physics, Weizmann Institute of Science, Rehovot 76100, Israel\\
$^4$Institut f\"ur Theoretische Physik, Freie
  Universit\"at Berlin, Arnimallee
     14, 14195 Berlin, Germany\footnote{Permanent address}}

\date{\today}
\begin{abstract}

Recent experiments on Coulomb drag in the quantum Hall regime have
yielded a number of surprises. The most
striking observations are that the Coulomb drag can become
negative in high Landau levels and that its temperature
dependence is non-monotonous. We develop a systematic diagrammatic
theory of Coulomb
drag in strong magnetic fields explaining these puzzling experiments.
The theory is applicable both in the diffusive and
the ballistic regimes; we focus on the experimentally relevant
ballistic regime (interlayer distance $a$ smaller than the cyclotron
radius $R_c$). It is shown that the drag at strong magnetic fields
is an interplay of two contributions
arising from different sources of particle-hole asymmetry, namely
the curvature of
the zero-field electron dispersion and the particle-hole asymmetry
associated with Landau quantization. The former contribution is
positive and governs the high-temperature increase in the
drag resistivity.
On the other hand,
the latter one, which is dominant at low $T$,
has an oscillatory sign
(depending on the difference in filling factors of the two layers)
and gives rise
to a sharp peak in the temperature dependence at $T$
of the order of the Landau level width.

\end{abstract}

\maketitle

\section{Introduction}

Coulomb drag between parallel two-dimensional electron
systems\cite{Eisenstein,Sivan} has developed into a powerful
probe of quantum-Hall
systems,\cite{Hill,Rubel,Gramila,Lilly98,kellogg02,kellogg03,Lok,Muraki}
providing information which is complementary to conventional
transport measurements. The drag signal is the voltage $V$ developing
in the open-circuit passive layer
when a current $I$ is applied in the active layer. The drag resistance
(also known as transresistance) is then defined by
$R_D=V/I$.
As a function of interlayer spacing $a$, the interlayer
coupling changes from weak at large
spacings where it can be treated in perturbation theory, to strong at
small spacings where it can result in
states with strong interlayer correlations \cite{kellogg02,kellogg03}.
In the present paper we will be concerned with the regime of weak
interlayer interaction.

In a simple picture of Coulomb drag, the carriers of the
active layer transfer momentum to the carriers of the passive layer by
interlayer
electron-electron scattering. Due to the open-circuit setup, a voltage
$V$ develops in the passive layer,
which balances this momentum transfer.
The phase space for interlayer scattering is proportional to the
temperature $T$
in either layer predicting a monotonous temperature dependence
$R_D\propto T^2$ of the  drag
resistance. Moreover, the signs of the voltages in active and passive
layer are
expected to be opposite (the same) for carriers of equal (opposite)
charge in the two
layers.\cite{MacDonald} It is conventional to refer to the sign
resulting for like (unlike) charges as positive
(negative) drag. It is worth emphasizing that, as the above
considerations imply, the non-zero value of drag in the regime of
weak interlayer interaction is entirely due to the violation of the
particle-hole symmetry.

Remarkably, experiments show that Coulomb drag behaves very
differently from these simple
expectations  when a perpendicular magnetic field $B$
is applied such that the Fermi energy $E_F$
is in a high Landau level, $E_F/\hbar\omega_c\gg 1$. ($\omega_c$ is
the cyclotron frequency.)
Several experiments\cite{Gramila,Lok} in the regime of weak interlayer coupling
observed negative drag when the filling factors in the two layers are
different. A more recent experiment\cite{Muraki} also reveals a
non-monotonic dependence on temperature.
While the drag resistivity shows a quadratic temperature dependence at
sufficiently
high temperatures, where drag is always positive, an additional peak
develops at low temperatures
which can have both a positive or a negative sign depending on the
filling-factor difference
between the two layers.

Early theoretical work\cite{Bonsager} on Coulomb drag in a magnetic
field in the limit of high Landau levels
showed that the magnetic field may strongly enhance the Coulomb
drag, as indeed observed experimentally. On the other hand, the
calculation of Ref.~\onlinecite{Bonsager}, as well as of a later
paper, \cite{Khaetskii} results in a strictly
positive transresistivity, in contradiction with the oscillatory sign
found in recent experiments.
As we discuss in detail below, a general formula
for the drag resistivity obtained in  Ref.~\onlinecite{Bonsager},
which looks like a natural generalization of the zero-$B$ result
\cite{MacDonald,Kamenev,Flensberg95} and also served
as a starting point for Ref.~\onlinecite{Khaetskii}, misses an
important contribution. This strongly restricts the range of
validity of the results of Refs.~\onlinecite{Bonsager,Khaetskii},
making them inapplicable under typical experimental conditions.
More recent work\cite{Oppen} showed that
Landau-level quantization can lead to sign changes in drag. However,
the results obtained in Ref.\ \onlinecite{Oppen}
suggested that unlike the experimental observation,
negative drag should be observed for equal filling factors in the two
layers. The temperature dependence of the drag resistivity was not
studied in \onlinecite{Oppen}.

In this paper, we present a systematic study of Coulomb
drag in the limit of high Landau levels.
We focus on the experimentally relevant limit of well-separated Landau
levels (LLs) in which the
LL broadening $\Delta$ is small compared to the LL spacing
$\hbar\omega_c$. Our starting point
is the diagrammatic Kubo formulation of Coulomb
drag\cite{Flensberg95,Kamenev} for weak interlayer
interaction. Disorder is included at the level of the self-consistent
Born approximation\cite{Ando}
(SCBA) which becomes exact in the limit of high Landau
levels \cite{Chalker}.

Our results are in good agreement with the experimental
observations. We find that at
high temperatures, the leading contribution to Coulomb drag is due to
the breaking of particle-hole
symmetry by the quadratic dispersion of the electrons. This
contribution which is analogous to the
conventional contribution to drag discussed above, always has a
positive sign and depends on
temperature as $T^2$. At temperatures $k_BT\ll\Delta$, we find that the
dominant contribution arises
from the breaking of particle-hole symmetry due to the Landau-level
structure. This contribution
gives rise to a peak in the temperature dependence and can take on
both positive and negative signs,
depending on the filling-factor difference of the two layers. In
particular, the sign is negative for equal filling
factors in the diffusive regime where the interlayer distance $a$ is
larger than the cyclotron radius $R_c$, as was found in Ref.\ 
\onlinecite{Oppen}.
We find, however, that this sign becomes negative in the
experimentally relevant ballistic regime ($a$ small compared to
$R_c$), in agreement with experiment.

This paper is organized as follows. Sec.\ \ref{background} briefly
summarizes the pertinent
background on the Kubo approach to Coulomb drag as well as on the
self-consistent Born approximation.
In Sec.\ \ref{triangle}, we present the diagrammatic calculation of
the triangle vertex entering
the expression for the drag conductivity, for well-separated LLs, both
in the diffusive
and in the ballistic regime of momenta. In Sec.\ \ref{screened}, we
collect the relevant
results for the screened interlayer interaction. These building blocks
are used in Sec.\
\ref{dragconductivity} to compute the drag resistivity. In this
section, we also compare our
results with experiment. Finally, Sec.\ \ref{summary} contains a
summary of our results and a discussion of prospects for future
research. In what follows, we set $\hbar=k_B=1.$

\section{Background}
\label{background}

\subsection{Drag}
\label{drag}

Our considerations are based on the Kubo approach to Coulomb drag
\cite{Kamenev,Flensberg95} which
expresses the drag conductivity $\sigma^{D}_{ij}({\bf Q},\Omega)$ in
terms of a current-current
correlation function,
\begin{equation}
\label{drag-conductivity-kubo}
   \sigma^{D}_{ij}({\bf Q},\Omega)={1\over \Omega S}\int_0^\infty dt\,
      e^{i\Omega t} \left\langle
      [j_i^{(1)\dagger}({\bf Q},t),j_j^{(2)}({\bf Q},0)]\right\rangle.
\end{equation}
where $i,j$ label the components of the drag conductivity tensor,
${\bf Q},\Omega$ denote the
wave vector and frequency of the applied field, $S$ is the area of the sample,
and $j_i^{(l)}$ denotes the $i$th component of the current operator in
the $l$th layer.
The $dc$ drag conductivity follows by taking the limit
\begin{equation}
\label{drag-conductivity-dc}
   \sigma^{D}_{ij}=\sigma^{D}_{ij}({\bf Q}=0,\Omega\to 0).
\end{equation}

\begin{figure}
\centering
\epsfxsize14.cm\epsfbox{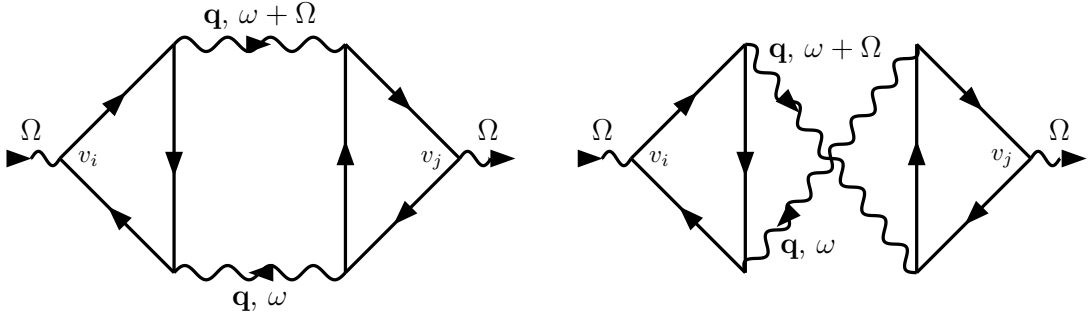}
\vskip .3truecm
\caption{ The diagrams contributing to the drag conductivity to leading
order in the interlayer interaction $U({\bf q},\omega)$ (wavy
lines). The full lines
represent the electron Green function. The external vertices labelled by the velocity operator 
$v_i$ are vector (current) vertices
while the internal vertices are scalar (density) vertices. }
\label{fig1}
\end{figure}

\begin{figure}
\centering
\epsfxsize7cm\epsfbox{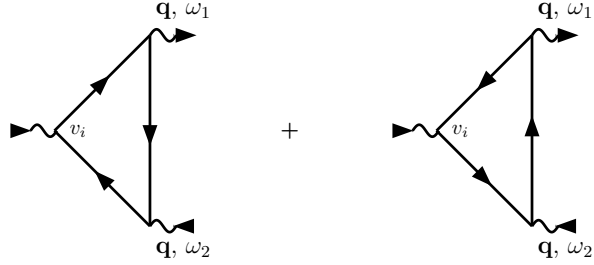} \vskip .3truecm
\caption{Diagrams defining the triangle vertex ${\bf \Gamma}({\bf
    q},\omega_1,\omega_2)$.
}
\label{fig2}
\end{figure}

When computing the retarded correlation function appearing in
Eq.\ (\ref{drag-conductivity-kubo}) within the Matsubara technique,
the leading diagrams
in the limit of weak (screened) interlayer interaction $U({\bf q},\omega)$
are shown in Fig.\ \ref{fig1}. Analytically, these diagrams are given
by the expression
\begin{equation}
\label{drag-conductivity-matsubara}
   \sigma_{ij}^{D}(i\Omega_k)={e^2T\over 2\Omega_k S}\sum_{{\bf q},\omega_n}
      \Gamma_i^{(1)}({\bf q},i\omega_n+i\Omega_k,i\omega_n)
      \Gamma_j^{(2)}({\bf q},i\omega_n,i\omega_n+i\Omega_k)
      U({\bf q},i\omega_n+i\Omega_k)U({\bf q},i\omega_n).
\end{equation}
Here, $\omega_n$ and $\Omega_k$ denote bosonic Matsubara frequencies and
the vector
${\bf\Gamma}^{(l)}({\bf q},i\omega_n,i\omega_m)$ is the triangle
vertex of layer $l$ as defined
by the diagrams in Fig.\ \ref{fig2}. Neglecting {\it intra}layer interactions,
it takes the analytical form
\begin{equation}
\label{triangle-matsubara}
   {\bf\Gamma}({\bf q},i\omega_n,i\omega_m)=
      T\sum_{\epsilon_k} {\rm tr} \left\{ {\cal G}(i\epsilon_k) e^{i{\bf
      qr}}{\cal G}(i\epsilon_k+i\omega_m){\bf v}
      {\cal G}(i\epsilon_k+i\omega_n)e^{-i{\bf qr}}+{\cal
      G}(i\epsilon_k)e^{-i{\bf qr}}{\cal G}(i\epsilon_k-i\omega_n)
      {\bf v}{\cal G}(i\epsilon_k-i\omega_m)e^{i{\bf qr}}\right \},
\end{equation}
where ${\cal G}$ denotes the Green function (for a particular
realization of the disorder potential),
$\epsilon_k$ is a fermionic Matsubara frequency, and ${\bf v}$
represents the velocity operator. The vertex $\bf\Gamma$ should be
averaged over realizations of disorder, as will be discussed in
Sec.~\ref{scba}.

Summing over the Matsubara frequency $\omega_n$, performing the
analytical continuation
to a real frequency $\Omega$, and finally taking the limit $\Omega\to 0$ yields
for the $dc$ drag conductivity \cite{Kamenev,Flensberg95}
\begin{equation}
\label{drag-conductivity-retarded}
   \sigma_{ij}^{D}={e^2\over 16\pi T S}\sum_{{\bf q}}\int_{-\infty}^\infty
     {d\omega\over \sinh^2(\omega/2T)}
      \Gamma_i^{(1)}({\bf q},\omega+i0,\omega-i0)
\Gamma_j^{(2)}({\bf q},\omega-i0,\omega+i0)|U({\bf q},\omega)|^2.
\end{equation}
In the sequel, we will use a short-hand notation,
${\bf\Gamma}({\bf  q},\omega)\equiv
{\bf\Gamma}({\bf  q},\omega+i0,\omega-i0)$. Note that the Onsager
relation $\sigma_{ij}^{12}(B)=\sigma_{ji}^{21}(-B)$ implies, in
combination with (\ref{drag-conductivity-retarded}), that
${\bf\Gamma}({\bf  q},\omega-i0,\omega+i0;B)=
{\bf\Gamma}({\bf  q},\omega+i0,\omega-i0;-B)$.

The experimentally measured drag resistivity can be expressed via
the drag conductivity as
\begin{equation}
\rho_{ij}^D=\rho_{ik}^{(1)}\ \sigma^D_{kl}\ \rho_{lj}^{(2)},\label{tensor-drag}
\end{equation}
where $\rho_{ik}^{(1,2)}$ are the resistivities of the layers.
Note that the minus sign corresponding to the standard tensor
inversion is absent
in this expression, according to the conventional definition of the
drag resistivity.
This definition yields a positive transresistivity in the absence of
a magnetic field.

The triangle vertex ${\bf \Gamma}({\bf q},\omega)$ is obtained by
analytic continuation of (\ref{triangle-matsubara}), see
Appendix~\ref{app:analytical} for detail. The result has the
form ${\bf \Gamma} =
{\bf \Gamma}^{(a)} + {\bf \Gamma}^{(b)}$ with the two contributions
\begin{eqnarray}
\label{triangle(a)}
  &&{\bf\Gamma}^{(a)}({\bf q},\omega)= \int {d\epsilon \over 4 \pi i}
    \tanh{\epsilon+\omega-\mu\over 2T} \nonumber\\
    &&\,\,\,\,\times{\rm tr}\left\{{\bf v}{\cal G}^+(\epsilon+\omega)
    e^{i{\bf qr}}{\cal G}^+(\epsilon)e^{-i{\bf qr}}{\cal
    G}^+(\epsilon+\omega)\right.
   -\left.{\bf v}{\cal G}^-(\epsilon+\omega)
    e^{i{\bf qr}}{\cal G}^-(\epsilon)e^{-i{\bf qr}}{\cal
    G}^-(\epsilon+\omega)\right\}
     +(\omega,{\bf q}\to-\omega,-{\bf q}),
   \\
\label{triangle(b)}
   &&{\bf\Gamma}^{(b)}({\bf q},\omega)=\int{d\epsilon\over 4 \pi i}
    (\tanh{\epsilon+\omega-\mu\over 2T}-\tanh{\epsilon-\mu\over 2T})
    \nonumber\\
    &&\,\,\,\,\times
    {\rm tr}\left\{{\bf v}{\cal G}^-(\epsilon+\omega)
    e^{i{\bf qr}}[{\cal G}^-(\epsilon)-{\cal G}^+(\epsilon)]e^{-i{\bf qr}}
    {\cal G}^+(\epsilon+\omega)\right\}
     +(\omega,{\bf q}\to-\omega,-{\bf q}).
\end{eqnarray}
Here, ${\cal G}^\pm(\epsilon)$ denotes the advanced/retarded Green
function and $\mu$ is the chemical potential.
Note that at zero magnetic field only ${\bf \Gamma}^{(b)}$
survives,\cite{Kamenev}
whereas ${\bf \Gamma}^{(a)}$ containing products of three advanced or
three retarded Green functions is zero. By contrast, in strong
$B$ both ${\bf \Gamma}^{(a)}$ and ${\bf \Gamma}^{(b)}$ should be
retained. Most importantly, we will show below that in the ballistic limit
there is a cancellation between ${\bf \Gamma}^{(a)}$ and ${\bf
  \Gamma}^{(b)}$ in the leading order.

For small $\omega$, the expressions for ${\bf\Gamma}({\bf q},\omega)$
simplify to
\begin{eqnarray}
    &&{\bf\Gamma}^{(a)}({\bf q},\omega)= {\omega \over 2 \pi i}
    {\rm tr}\left\{{\bf v}{\cal G}^+(\epsilon)
    e^{i{\bf qr}}{\cal G}^+(\epsilon)e^{-i{\bf qr}}{\cal G}^+(\epsilon)
     -({\cal G}^+\to {\cal G}^-)\right\}
    \label{gammaa}\\
   &&{\bf\Gamma}^{(b)}({\bf q},\omega)={\omega\over i\pi }
    {\rm tr}\left\{{\bf v}{\cal G}^-(\epsilon)
    e^{i{\bf qr}}[{\cal G}^-(\epsilon)-{\cal G}^+(\epsilon)]e^{-i{\bf qr}}
    {\cal G}^+(\epsilon)\right\}. \label{gammab}
\end{eqnarray}
For well-separated LLs, this approximation holds as long as $\omega$
is small compared
to the width $\Delta$ of the LL.
It is also useful to note that ${\bf \Gamma}^{(a)}({\bf q},\omega)$
can be expressed as
\begin{equation}
\label{gammaa-short}
    {\bf\Gamma}^{(a)}({\bf q},\omega)= {\omega \over \pi }
    \nabla_{\bf q}{\rm Im}\,{\rm tr}\left\{ e^{i{\bf qr}}{\cal
    G}^+(\epsilon)e^{-i{\bf qr}}
     {\cal G}^+(\epsilon)\right\},
\end{equation}
which shows that ${\bf\Gamma}^{(a)}({\bf q},\omega)$ gives only a
longitudinal contribution
(parallel to ${\bf q}$) to ${\bf\Gamma}({\bf q},\omega)$.

\subsection{Impurity diagram technique in high Landau levels -- SCBA}
\label{scba}

In this subsection, we discuss the averaging over the random potential
of impurities. We assume white-noise disorder,
characterized by zero mean,
$\langle U({\bf r})\rangle = 0$, and by the correlation function
$$\langle U({\bf r})U({\bf r}')\rangle =
{1\over 2\pi \nu_0 \tau_0}\delta({\bf r}-{\bf r}'),$$
where $\nu_0=m/2\pi$ denotes the zero-B density of states per spin
and $\tau_0$ the zero-B elastic scattering time.
We perform the averaging in the self-consistent
Born approximation (SCBA).
This approximation, which neglects diagrams with
crossing impurity lines, can be shown to give the leading
contribution when the Fermi energy $E_F$ is in a high LL with LL index
$N\gg 1$.\cite{Chalker}
Strictly speaking, the disorder potential in the experimental samples
is expected to
be correlated on the scale of the distance of the
two-dimensional electron layer
from the donor layer.
However, we find that the experimental observation can already be understood
when considering white-noise disorder and that a finite correlation
length of disorder
does not qualitatively change our conclusions.\cite{unpublished}

Within the SCBA for well-separated Landau levels,\cite{Ando} the
impurity average of the Green function, denoted
by $G^\pm(\epsilon)$, is diagonal in the LL basis $|nk\rangle$ in the
Landau gauge and takes the expression
\begin{equation}
\label{scba-green}
    G_{n}^\pm(\epsilon) = {1\over \epsilon - E_n -\Sigma^\pm(\epsilon)}
\end{equation}
with the LL energies $E_n=\omega_c(n+1/2)$.
For energies $\epsilon$ within a Landau level, the self-energy is given by
\begin{equation}
\label{scba-sigma}
\Sigma_n^\pm (\epsilon)
= {1\over 2} \{ \epsilon -E_n \pm i [ \Delta^2
-(\epsilon-E_n)^2]^{1/2}\}.
\end{equation}
Here, the LL index $n$
is chosen such that $|\epsilon -E_n| < \Delta$. The LL broadening
$\Delta$ can be expressed
in terms of the zero-field scattering time $\tau_0$ as
\begin{equation}
\label{scba-gamma}
\Delta^2=2\omega_c/\pi \tau_0.
\end{equation}
The density of states is
\begin{equation}
\label{scba-nu}
\nu(\epsilon) = 1/\pi^2\ell^2\Delta^2\tau(\epsilon)=\nu_0 \tau_0
[\Delta^2 - (\epsilon-E_n)^2]^{1/2}
\end{equation}
with the elastic scattering time
\begin{equation}
\label{scba-tau}
\tau(\epsilon) = [\Delta^2 - (\epsilon-E_n)^2]^{-1/2}.
\end{equation}
Here, $\ell=(1/eB)^{1/2}$ denotes the magnetic length.

\begin{figure}
\centering
\epsfxsize10cm\epsfbox{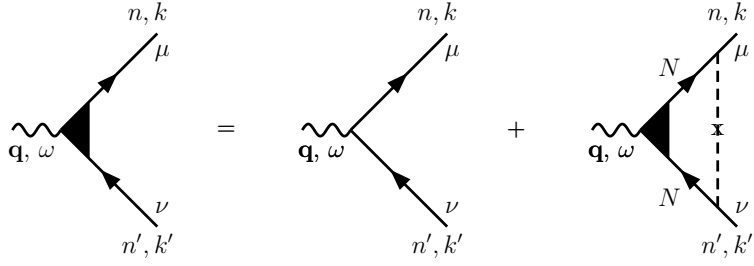} \vskip .3truecm
\caption{ Diagrammatic representation of the equation for the vertex
  corrections
$\gamma^{\mu\nu}_{nk,n'k'} (\epsilon+\omega,\epsilon;{\bf q})$ (full triangle at vertex)
of the scalar (density) vertices in the SCBA. Dashed lines represent
impurity scattering. We also indicate that for well-separated LLs, the internal Green functions in 
the right-most diagram should be evaluated in the valence LL $N$ which can differ from the LL labels $n,n'$
of the external Green functions.
}
\label{fig3}
\end{figure}

In principle, disorder leads to vertex corrections of both the vector
and the scalar vertices
of the triangle diagram ${\bf \Gamma}$. However, for white-noise
disorder there are no vertex
corrections of the vector vertex. The vertex corrections of the scalar
vertices generally involve
impurity ladders (cf.\ Fig.\ \ref{fig3}) and
turn out to be independent of the LL indices $n$ and
$n'$,\cite{bonsager-comment}
\begin{equation}
\label{scba-vertex-def}
   \gamma^{\mu\nu}_{nk,n'k'} (\epsilon+\omega,\epsilon;{\bf q}) =
       \gamma^{\mu\nu}({\bf q},\omega)\,
       \langle n k |e^{i{\bf qr}}| n' k'\rangle.
\end{equation}
Here, the indices $\mu,\nu =\pm$ indicate the type of Green functions
involved in the vertex.
In the limit of well-separated Landau levels, one finds the explicit
expressions
for the vertex corrections at $\omega=0$
\begin{eqnarray}
\label{scba-vertex++}
  \gamma^{++}({\bf q},\omega)&=&{1\over 1 - J_0^2(qR_c)[\Delta/2\Sigma^-]^2} \\
  \label{scba-vertex+-}
  \gamma^{+-}({\bf q},\omega)&=&{1\over 1 - J_0^2(qR_c)}
\end{eqnarray}
where $J_n(z)$ denotes the Bessel functions.
The derivation of these expressions is reviewed in Appendix~\ref{app:vertex}.

For later reference, we also collect relevant matrix elements between
LL eigenstates
$|nk\rangle$ in the Landau gauge ${\bf A}=B(0,x)$. The vector vertex
involves the matrix elements
\begin{eqnarray}
\label{matrix-element-vx}
   \langle nk|v_x|n'k'\rangle&=&\delta_{kk'}{i\over m\ell \sqrt{2}}\left\{
     \sqrt{n}\delta_{n,n'+1}-\sqrt{n+1}\delta_{n,n'-1}\right\}\\
\label{matrix-element-vy}
   \langle nk|v_y|n'k'\rangle&=&\delta_{kk'}{1\over m\ell \sqrt{2}}\left\{
     \sqrt{n}\delta_{n,n'+1}+\sqrt{n+1}\delta_{n,n'-1}\right\}.
\end{eqnarray}
In the limit of high Landau levels, $n\sim N\gg 1$, one can use
quasiclassical approximations for
these matrix elements, namely $\langle nk|v_x|n\pm 1 k'\rangle\simeq
\mp i\delta_{kk'}v_F/ 2 $ and
$ \langle nk|v_y|n\pm 1 k'\rangle\simeq \delta_{kk'}v_F/ 2 $,
with the Fermi velocity $v_F$.

The scalar vertex involves the matrix element
\begin{eqnarray}
\label{matrix-element-density}
   \langle nk|e^{i{\bf qr}}|n'k'\rangle = \delta_{q_y,k-k'}
{2^{n'-n}n'!\over n!}\exp[-{1\over 4}q^2\ell^2-{i\over
  2}q_x(k+k')\ell^2]
[(q_y+iq_x)\ell]^{n-n'} L_{n'}^{n-n'}(q^2\ell^2/2)\qquad (n\ge n'),
\end{eqnarray}
where $L_m^n$ is the associated Laguerre polynomial. The expression
for the matrix element for $n<n'$ can be obtained from
(\ref{matrix-element-density}) by complex conjugation with the
replacement $q\to -q$, $nk\leftrightarrow n'k'$.
Since the characteristic LL indices are large, $n,n'\gg 1, |n-n'|$, and
relevant momenta are small compared to the Fermi momentum, $q\ll k_F$,
Eq.~(\ref{matrix-element-density}) can be simplified by using the
quasiclassical approximation,
\begin{equation}
\label{matr-el-quasicl}
   \langle nk| e^{i{\bf qr}}| n'k'\rangle \simeq
\delta_{q_y,k-k'} i^{n-n'}
   e^{-i\phi_q(n-n')}e^{-i q_x(k+k')\ell^2/2} J_{n-n'}(qR_c^{(m)}),
\end{equation}
where $\phi_q$ is the polar angle of ${\bf q}$,
$R_c^{(n)}=\ell\sqrt{2n+1}$ is the cyclotron radius of the $n$-th LL,
and $m=(n+n')/2$. For most of the calculations below, the
dependence of the cyclotron radius on the LL index in the vicinity of
the Fermi level will be immaterial, and we will drop the corresponding
superscript and simply write $R_c$. The $n$-dependence of $R_c$ will,
however, be crucial for the evaluation of the contribution to the drag
related to the curvature of the electron spectrum, see
Sec.~\ref{conventional}.
In view of the rotational invariance of ${\bf \Gamma}({\bf
  q},\omega)$, it is sufficient to calculate it for a certain direction
of the wave vector ${\bf q}$. Choosing ${\bf q}$ to point along the
positive $x$-axis, we simplify (\ref{matr-el-quasicl}) to the form
\begin{equation}
\label{matr-el-x-quasicl}
   \langle nk| e^{iqx}| n'k'\rangle \simeq \delta_{kk'} i^{n-n'}
   e^{-iqk\ell^2} J_{n-n'} (qR_c).
\end{equation}

\section{Triangle vertex ${\bf \Gamma}({\bf q},\omega)$}
\label{triangle}

\subsection{Leading order}
\label{leading}

\begin{figure}
\centering
\epsfxsize8cm\epsfbox{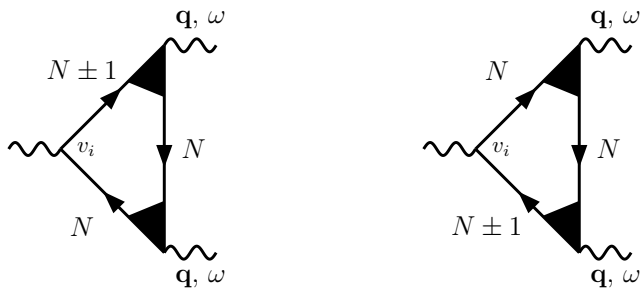} \vskip .3truecm
\caption{
The diagrams contributing to the triangle vertex in SCBA to leading
order in the limits of well-separated
Landau levels $\Delta/\omega_c$ and large $N$.
}
\label{fig4}
\end{figure}

We now turn to an evaluation of the disorder-averaged triangle vertex
${\bf \Gamma}({\bf q},\omega)$ for well
separated Landau levels, $\Delta/\omega_c\ll 1,$ in the limit in
which the Fermi energy is in
a high Landau level, $N\gg 1$. The relevant diagrams are shown in
Fig.\ \ref{fig4}. We begin by considering the low-temperature
limit, $\omega,T\ll \Delta.$

In the limit under consideration, the calculation is simplified as
follows. Very generally for white-noise   disorder, there are no
vertex corrections of the vector vertex. By contrast, vertex
corrections of the  scalar vertices have to be retained. To leading
order in $\Delta/\omega_c$, two of the three  Green functions in
Eqs.\ (\ref{gammaa}) and (\ref{gammab}) should be evaluated in the
$N$th Landau level  in which the Fermi energy is situated. Since the
velocity operator has matrix elements only between  states in
neighboring Landau levels, one of the Green functions adjacent to the
vector vertex must be  taken in Landau levels $N\pm 1$. This is
illustrated in Fig.\ \ref{fig4}.

We first consider the contribution
${\bf \Gamma}^{(a)}({\bf q},\omega)$. In this case, it is most convenient
to start from the simplified expression in Eq.\ (\ref{gammaa-short})
  in which to leading order in
$\Delta/\omega_c$, both remaining Green functions can be
  evaluated in Landau level $N$. Using the
identity $\nabla_{\bf q} [J_0(qR_c)]^2 = -2 {\bf \hat q} R_c
  J_0(qR_c)J_1(qR_c)$ (with ${\bf \hat q}
={\bf q}/q$), one obtains
\begin{equation}
\label{leading-gammaa}
     {\bf \Gamma}^{(a)}({\bf q},\omega) = - 2 {\hat{ \bf q}} {\omega R_c
   \over \pi^2\ell^2}\,
   J_0(qR_c) J_1(qR_c) \, {1\over 2i}\{ [G^+_N\gamma^{++}]^2
   - [G^-_N\gamma^{--}]^2 \},
\end{equation}
where $\gamma^{\mu\nu}\equiv \gamma^{\mu\nu}({\bf q},0)$ and
the factor 2 accounts for the spin degeneracy.
The calculation for ${\bf \Gamma}^{(b)}({\bf q},\omega)$ in Eq.\
(\ref{gammab}) yields
\begin{equation}
\label{leading-gammab}
   {\bf \Gamma}^{(b)} ({\bf q},\omega) =  2 {\bf {\hat q}}{\omega
   R_c\over \pi^2\ell^2}\,
   J_0(qR_c) J_1(qR_c)\, {1\over 2i} [G^+_N\gamma^{++}\gamma^{+-} -
   G_N^- \gamma^{+-} \gamma^{--}]\
   [G_N^+ + G_N^-] .
\end{equation}
Summing both contributions, one obtains
\begin{equation}
\label{leading-gamma}
   {\bf \Gamma} ({\bf q},\omega) =  {\bf {\hat q}}\ {4 \omega R_c\over
   \pi^2\ell^2}\,
   J_0(qR_c) J_1(qR_c)\,  {\rm Re}\, [G_N^+(
   \gamma^{++} -\gamma^{+-})]\  {\rm Im}\, [G^+_N\gamma^{++}].
\end{equation}
For arbitrary $T,\omega<\omega_c$, this contribution takes the form~\cite{foot-vert}
\begin{eqnarray}
\label{leading-gamma-finiteTw}
   {\bf \Gamma} ({\bf q},\omega) &=&  {\bf {\hat q}}\ {8 R_c\over  \pi^2\ell^2 \Delta^2}\,
   {J_1(qR_c)\over  J_0(qR_c)}\,
\int_{-\infty}^\infty d\epsilon \
\left[{\rm tanh}{\epsilon+\omega-\mu\over 2 T}- {\rm tanh}{\epsilon-\mu\over 2 T}\right]\nonumber \\
&\times&
{\rm Re}\, [ \gamma^{-+}({\bf q},\omega)-\gamma^{++}({\bf q},\omega)]\
{\rm Im}\, \gamma^{++}({\bf q},\omega).
\end{eqnarray}

Since the interlayer interaction is suppressed
at large momenta $q$ by a factor $e^{-qa}$, where $a$ is the
interlayer distance, the drag conductivity
(\ref{drag-conductivity-retarded}) is governed by momenta  $q<1/a$.
Depending on the relation between $a$ and the cyclotron radius $R_c$,
one distinguishes between the diffusive ($a\gg R_c$) and the ballistic
($a\ll R_c$) regimes. While in the former case, only ``diffusive''
momenta ($qR_c\ll 1$) are relevant, in the latter case both
``ballistic'' ($qR_c\gg 1$) and diffusive momenta contribute to the
drag conductivity (\ref{drag-conductivity-retarded}). Experimentally,
when the transresistivity is measured in moderately strong magnetic
fields (i.e.\ in high Landau levels), the condition $R_c>a$ is
typically satisfied. For this reason, we mainly concentrate on the
ballistic regime in this paper. In Secs.~\ref{diffusive} and
\ref{ballistic} we will calculate the triangle vertex ${\bf
  \Gamma}({\bf q},\omega)$
in the diffusive and ballistic ranges of momenta, respectively. These
results will be used in Sec.~\ref{dragconductivity} for the calculation of
the drag resistivity.

\subsection{Diffusive momenta, $qR_c\ll 1$}
\label{diffusive}

In the diffusive range of momenta, $qR_c\ll 1$, we can expand the
Bessel functions in
the expressions for the vertex corrections, Eqs.\
(\ref{scba-vertex++}) and (\ref{scba-vertex+-}).
Due to the singular behavior of the vertex correction $\gamma^{+-}$ at small
momenta $q$, we have $\gamma^{+-}\gg \gamma^{++},\gamma^{--}$, so that
only the contribution proportional to $\gamma^{+-}$ should be retained
in (\ref{leading-gamma}). This yields
\begin{equation}
\label{leading-diffusive}
   {\bf \Gamma} ({\bf q},\omega) =  - {\bf {\hat q}}{4 \omega
   R_c(qR_c)\over \pi^2\ell^2\Delta^2}\,
   {2\over (qR_c)^2} {\mu-E_N\over [\Delta^2-(\mu-E_N)^2]^{1/2}}.
\end{equation}
More generally,
at small momenta one should also take into account the frequency dependence
of $\gamma^{+-}$, which has the structure of a diffusion pole,
\begin{equation}
  \label{gamma+-diff}
\gamma^{+-}({\bf q},\omega)={1\over \tau(\epsilon)[D(\epsilon)q^2-i\omega]},
\end{equation}
where $D(\epsilon)=R_c^2/2\tau(\epsilon)$ is the (energy-dependent) diffusion
constant in a strong magnetic field.
Eq.\ (\ref{leading-diffusive}) is then generalized to
\begin{equation}
\label{diffusive-omega}
   {\bf \Gamma} ({\bf q},\omega) =  - {\bf {\hat q}}\
   {4 \omega qR_c^2\over \pi^2\ell^2\Delta^2}\,
 \frac{D(\mu)
  q^2}{[D(\mu) q^2]^2+\omega^2}\
  (\mu-E_N).
\end{equation}
This result can be recast in the form ($n_e$ is the electron concentration)
\begin{equation}
\label{diffusive-oppen}
    e{\Gamma}_i({\bf q},\omega) =
2{d\sigma_{ij} \over
    d(en_e)}\cdot{q}_j\ {\rm Im} \Pi ({\bf q},\omega) \simeq
2{d\sigma_{xx} \over
    d(en_e)}\cdot{q}_i\ {\rm Im} \Pi ({\bf q},\omega),
\end{equation}
which allows for a simple interpretation as a nonlinear
susceptibility.\cite{Aleiner-drag,Oppen}
This rewriting of Eq.\ (\ref{diffusive-omega}) uses the result\cite{Ando}
\begin{eqnarray}
\label{scba-sigmaxx}
   \sigma_{xx} = {e^2\over \pi^2} N \left[1-{(\mu-E_N)^2\over
   \Delta^2}\right]
\end{eqnarray}
for the diagonal conductivity in SCBA ($d\sigma_{xx}/d n_e \gg d\sigma_{xy}/d n_e$
for separated LLs) and
\begin{equation}
\label{diffusive-polarization}
   \Pi({\bf q},\omega) = 2\nu(\mu) {D(\mu)q^2 \over D(\mu) q^2 -i\omega},
\end{equation}
for the polarization operator in the diffusive limit.
It is worth emphasizing
that the diffusive result in Eq.\ (\ref{diffusive-omega})
arises from ${\bf\Gamma}^{(b)}$ only,
since the other contribution ${\bf\Gamma}^{(a)}$ does not contain the
vertex correction $\gamma^{+-}$. Note that the authors of
Ref.~\onlinecite{Bonsager} failed to obtain the leading diffusive
contribution (\ref{diffusive-omega}), (\ref{diffusive-oppen}), because
of an incorrect treatment of vertex corrections~\cite{bonsager-comment,foot-vert}.
For the same reason,
they missed the ${\cal O}(1/qR_c)$ contribution
[Eq.~(\ref{gamma-1/qrc}) below] which becomes
important in the ballistic regime, as we are going to discuss.

\subsection{Ballistic momenta, $qR_c\gg 1$}
\label{ballistic}

\subsubsection{Cancellation of leading contribution and
the ${\cal O}(1/qR_c)$ contribution   from vertex
corrections}
\label{cancellation}

In the leading order in the ballistic range of momenta,
$qR_c\gg 1$, the vertex corrections in
Eqs.\ (\ref{scba-vertex++}) and (\ref{scba-vertex+-}) can be neglected,
\begin{equation}
\label{vertex-ballistic}
   \gamma^{++}({\bf q},\omega)\simeq \gamma^{+-}({\bf q},\omega)\simeq 1.
\end{equation}
Inserting this into
Eq.\ (\ref{leading-gamma}) for the triangle
vertex ${\bf \Gamma}({\bf q},\omega)$, we immediately see that the
triangle vertex vanishes to this order. We emphasize that
${\bf \Gamma}^{(a)}$ and ${\bf \Gamma}^{(b)}$ do not vanish separately but rather
cancel each other in the leading order.

Thus,
to obtain a non-zero answer for ${\bf \Gamma}({\bf q},\omega)$
from Eq.\ (\ref{leading-gamma}), we need to consider the vertex corrections
in Eqs.\ (\ref{scba-vertex++}) as (\ref{scba-vertex+-}) in
next-to-leading order in $qR_c$.
In this way, one finds from Eq.\ (\ref{leading-gamma})
\begin{equation}
\label{gamma-1/qrc}
   {\bf \Gamma}^{(1/qR_c)} ({\bf q},\omega) =  - {\bf {\hat
   q}}{64 \omega R_c\over \pi^2\ell^2}\,
   {(\mu-E_N)[\Delta^2 - (\mu-E_N)^2]^{3/2}\over \Delta^6}
   J_1(qR_c) J^3_0(qR_c).
\end{equation}
Here, we introduced a superscript on ${\bf \Gamma}({\bf q},\omega)$ in
order to distinguish
this contribution from other contributions computed below.
At finite $T$ and $\omega$ (assuming $T,\omega<\omega_c$),
we find
\begin{eqnarray}
\label{gamma-1/qrc-finitTw}
   {\bf \Gamma}^{(1/qR_c)} ({\bf q},\omega) &=&
   - {\bf {\hat q}}{16 \omega R_c\over \pi^2\ell^2 \Delta^2}\,J_1(qR_c) J^3_0(qR_c)
      \int_{-\infty}^{\infty}d\epsilon
  \left(\tanh\frac{\epsilon+\omega/2-\mu}{2T}
- \tanh\frac{\epsilon-\omega/2-\mu}{2T}
\right)
\nonumber \\
&\times&
{\rm Re}\left[1-{(\epsilon+\omega/2-E_N)^2\over\Delta^2}\right]^{1/2}
{\rm Re}\left[1-{(\epsilon+\omega/2-E_N)^2\over\Delta^2}\right]^{1/2}
\nonumber \\
&\times&
\left\{
 \frac{\epsilon+\omega/2-E_N}{\Delta}
  \left[1-{(\epsilon-\omega/2-E_N)^2\over\Delta^2}\right]
  +
\frac{\epsilon-\omega/2-E_N}{\Delta}
  \left[1-{(\epsilon+\omega/2-E_N)^2\over\Delta^2}\right]
  \right\}.
\end{eqnarray}

The contribution (\ref{gamma-1/qrc}) to ${\bf \Gamma}({\bf q},\omega)$
has been obtained in
leading order in the limit $\Delta/\omega_c\ll 1$ and $q/k_F\ll 1$
and in next-to-leading order
in $qR_c\gg 1$. Thus, we are also forced to consider separately
next-to-leading order corrections
in the parameters $\Delta/\omega_c\ll 1$ and $q/k_F\ll 1$, with the
other two parameters kept
in leading order.

Before we turn to these calculations, we briefly remark that
the leading-order cancellation in the ballistic regime was missed in
Ref.\ \onlinecite{Oppen} since the contribution from ${\bf
  \Gamma}^{(a)}$ was overlooked.
The results obtained there for the diffusive regime remain valid since
in this case ${\bf \Gamma}^{(a)}$ is negligible compared to
${\bf \Gamma}^{(b)}$, see Sec.~\ref{diffusive}.

\subsubsection{Contributions of order $\Delta/\omega_c$}
\label{llmixing}

In this section, we consider the first corrections to the leading order
in $\Delta/\omega_c$ to the triangle vertex ${\bf \Gamma}({\bf
  q},\omega)$,
while working to leading order in the ballistic limit $qR_c\gg 1$ for
high Landau levels $N\gg 1$.
While such corrections are of higher order in the small parameter
$\Delta/\omega_c$, this
smallness may be compensated by a large factor $qR_c$ since it turns
out that in this case there is no
cancellation between ${\bf \Gamma}^{(a)}$ and ${\bf \Gamma}^{(b)}$.

\begin{figure}
\centering
\epsfxsize8cm\epsfbox{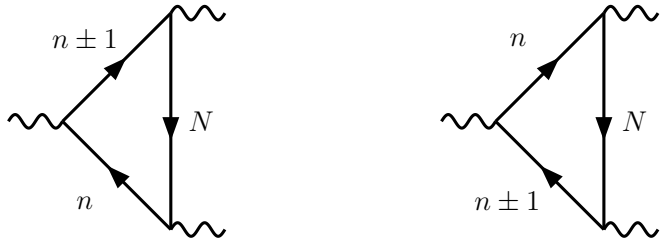} \vskip .3truecm
\caption{Diagram contributing to corrections of order
  $\Delta/\omega_c$ to the triangle vertex.
Here both Green function adjacent to the vector vertex should be
evaluated in Landau levels different
from $N$.}
\label{fig5}
\end{figure}

Corrections of order $\Delta/\omega_c$ arise from two sources:
(i) The Green functions adjacent to
the current vertex are {\it both} evaluated in Landau levels different from $N$. 
(Note that the Green function
between the scalar vertices must still be evaluated in the $N$th
Landau level because
$G^+_{n}-G^-_n\sim \Delta/\omega_c^2$ for $n\neq N$.) This
contribution is depicted in
Fig.\ \ref{fig5}. (ii) The diagrams in Fig.\ \ref{fig4} can be
evaluated more accurately, keeping
corrections in $\Delta/\omega_c$, which arise from keeping the self-energy
parts of the Green functions
of Landau levels $N\pm 1$. Note that we may now neglect vertex corrections at the
scalar vertices because we consider the leading order in  $qR_c\gg 1$.

Details of this calculation are presented in Appendix~\ref{app:gamma}. Here we only state
the results. The contribution (i) vanishes for both ${\bf
  \Gamma}^{(a)}$ and ${\bf \Gamma}^{(b)}$.
The contribution (ii) turns out to still give a vanishing contribution
to the longitudinal triangle vertex, due to the cancellation between
  ${\bf \Gamma}^{(a)}$ and
${\bf \Gamma}^{(b)}$ described above. However, the transverse contribution to
${\bf\Gamma}^{(b)}$ no longer vanishes when considering corrections in
$\Delta/\omega_c$. In this way, we obtain the contribution
\begin{equation}
\label{gamma-llmixing}
   {\bf \Gamma}^{(\Delta/\omega_c)}({\bf q},\omega)=-{\bf \hat
   q}\times {\bf \hat z}
   {16 \omega R_c\over\pi^2\ell^2} {(\mu-E_N)[\Delta^2-(\mu-E_N)^2]\over
   \omega_c \Delta^4} J_0(qR_c)J_1(qR_c)
\end{equation}
to the triangle vertex.

As we will see below, the  $\Delta/\omega_c$
contribution is of crucial importance for understanding the
experimental findings. We mention that
this term was lost in Ref.~\onlinecite{Bonsager} (in addition to the
$1/qR_c$ contribution missing there because of an inaccurate treatment of
vertex corrections) in the course of the so-called
``triangles-to-bubbles'' transformation.
Specifically, in Ref.~\onlinecite{Bonsager} the
self-energy in the Green functions connected by the current vertex
was neglected compared to the cyclotron frequency, which
obviously misses corrections of order of $\Delta/\omega_c$.

Eq.\ (\ref{gamma-llmixing}) is derived in the low-temperature limit,
when $T,\omega\ll\Delta$.  To analyze the temperature dependence of
the drag, we will need the $\Delta/\omega_c$-contribution also at
higher temperatures. We find for arbitrary relations between $T$ and $\Delta$
and $T,\omega<\omega_c$ that
\begin{eqnarray}
{\bf \Gamma}^{(\Delta/\omega_c)}({\bf q},\omega)&=&-{\bf \hat
  q}\times {\bf \hat z} \frac{8 R_c}{\pi^2 \ell^2 \Delta}
J_1(q R_c) J_0(q R_c) \nonumber \\
&\times&\int_{-\infty}^{\infty}d\epsilon
  \left(\tanh\frac{\epsilon+\omega/2-\mu}{2T}
- \tanh\frac{\epsilon-\omega/2-\mu}{2T}
\right)
\nonumber \\
&\times&
  \frac{\epsilon-E_N}{\Delta}\ {\rm Re} \left[1-{(\epsilon+\omega/2-E_N)^2\over
  \Delta^2}\right]^{1/2}
{\rm Re} \left[1-{(\epsilon-\omega/2-E_N)^2\over \Delta^2}\right]^{1/2}.
\label{finite-w-2}
\end{eqnarray}

\subsubsection{The conventional contribution of order $q/k_F$}
\label{conventional}

In this section, we compute the contribution to ${\bf \Gamma}$ due to
terms of order $q/k_F$ relative
to the leading order. Such terms arise from a more accurate treatment
of the matrix elements involved
in the scalar vertices, for which we now use the more accurate expressions
\begin{eqnarray}
   \langle N\pm 1| e^{i{\bf qr}}| N\rangle \langle N | e^{-i{\bf qr}}|
       N\rangle
       &\simeq& i J_1(qR_c[1\pm {1\over 4N}]) J_0(qR_c) \\
   \langle N| e^{i{\bf qr}}| N\rangle \langle N | e^{-i{\bf qr}}| N\pm
       1\rangle
      &\simeq& i J_1(qR_c[1\pm {1\over 4N}]) J_0(qR_c)
\end{eqnarray}
together with
\begin{equation}
    J_1(qR_c[1\pm {1\over 4N}]) \simeq J_1(qR_c) \pm {q\over 2k_F} J_0(qR_c).
\end{equation}
Thus, such terms give rise to a contribution of the order of $q/k_F$
relative to the  naive leading order [which vanishes because of the
cancellation between ${\bf \Gamma}^{(a)}$ and ${\bf
  \Gamma}^{(b)}$].

We find that such corrections arise only for the contribution ${\bf
  \Gamma}^{(b)}$, yielding
\begin{equation}
   {\bf \Gamma}^{(q/k_F)}({\bf q},\omega)=
   {\bf q}\times \hat{\bf z}\,{8 \omega \over \pi^2}\,{\Delta^2 - (\mu-E_N)^2 \over
     \Delta^4} J^2_0(qR_c).
\end{equation}
Similarly to Eq.~(\ref{finite-w-2}),
we  generalize this ${\cal O}(q/k_F)$ contribution
to the finite-$T$ case,
\begin{eqnarray}
 {\bf \Gamma}^{(q/k_F)}({\bf q},\omega)&=&{\bf q}\times \hat{\bf z}
  \,\frac{4 J_0^2(q R_c) }{\pi^2 \Delta^2}
\int_{-\infty}^{\infty}d\epsilon \left(\tanh\frac{\epsilon+\omega/2-\mu}{2T}
- \tanh\frac{\epsilon-\omega/2-\mu}{2T}\right)\nonumber \\
&\times&
{\rm Re}\ \left[1-{(\epsilon+\omega/2-E_N)^2\over \Delta^2}\right]^{1/2}
{\rm Re}\ \left[1-{(\epsilon-\omega/2-E_N)^2\over \Delta^2}\right]^{1/2}.
\label{finite-w-3}
\end{eqnarray}

This expression can also be rewritten as
\begin{equation}
\label{gamma-qkf}
  e {\bf \Gamma}^{(q/k_F)}({\bf q},\omega)= {\bf q}\times \hat{\bf z} \,{2\sigma_{xy} \over e n_e}
\,
    {\rm Im}\, \Pi({\bf q},\omega)
\end{equation}
with the polarization operator $\Pi({\bf q},\omega)$ for the ballistic regime
[cf.\ Eq.\ (\ref{pi-qw-ballistic-q}) below] and the Hall
conductivity
\begin{equation}
\sigma_{xy}={ en_e\over B} - {e^2\over \pi^2}\ N\ {\Delta\over
  \omega_c}\left[1-{(\mu-E_N)^2\over \Delta^2}\right]^{3/2}
\label{sigmaxy-SCBA}
\end{equation}
in SCBA. It can be checked that Eq.~(\ref{gamma-qkf}) is valid for arbitrary $T$,
including $T\agt \omega_c$.

The $q/k_F$ contribution arises from taking into account the
dependence of the
cyclotron radius and hence the
velocity on the Landau level number, which is a
direct consequence of the curvature of the zero-$B$
electron spectrum. It is thus natural that the obtained result
(\ref{gamma-qkf}) is a high-magnetic field analog of the
conventional contribution to ${\bf \Gamma}$.\cite{MacDonald}
Only this contribution was retained in
Refs.~\onlinecite{Bonsager,Khaetskii}, while the other contributions related
to the particle-hole asymmetry due to the LL quantization were lost
there.

\section{Screened interlayer interaction}
\label{screened}

In this section, we summarize the results for the screened
interlayer interaction\cite{Kamenev}
\begin{equation}
   U_{12}({\bf q},\omega) = {V_{12}(q)\over [1+V({\bf q})\Pi_1({\bf
   q},\omega)][1+V({\bf q})\Pi_2({\bf q},\omega)]
   -V^2_{12}({\bf q})\Pi_1({\bf q},\omega)\Pi_2({\bf q},\omega)}.
   \label{U12-general}
\end{equation}
Here, $V({\bf q})=2\pi e^2/q$ denotes the bare {\it intra}layer
interaction and $V_{12}({\bf q})=V({\bf q})
e^{-q a}$ is the bare {\it inter}layer interaction, $a$ denotes the
distance between the layers.
The polarization operator of layer $l$ is denoted by $\Pi_l({\bf q},\omega)$.
For $q$ small compared to the Thomas-Fermi screening wave vectors
$\kappa_{0,l}=4\pi e^2\nu_{0,l}$
($l=1,2$ labels the layer and $\nu_{0,l}$ denotes the zero-field
density of states per spin of
layer $l$), this can be approximated as
\begin{equation}
   U_{12}({\bf q},\omega) \simeq {\pi e^2 q\over \kappa_{0,1}\kappa_{0,2}
     \sinh(q a)}
     {2\nu_{0,1}\over \Pi_1({\bf q},\omega)}{2\nu_{0,2}\over
     \Pi_2({\bf q},\omega)}.
     \label{U12-simple}
\end{equation}

In the
random-phase approximation, the polarization operator in a strong magnetic
field has the form
\begin{eqnarray}
\label{pi-qw}
\Pi({\bf q},\omega)&=&{1\over \pi \ell^2}\sum_{n,m}J^2_{n-m}(q R_c)
\int_{-\infty}^\infty {d\epsilon\over 2\pi i}\ n_F(\epsilon)
\left\{G_n^+(\epsilon+\omega) \
[G_m^+(\epsilon)\gamma^{++}({\bf q},\omega)- G_m^-(\epsilon)\gamma^{+-}({\bf q},\omega)]
\right.
\nonumber \\
&+&  \left. G_n^-(\epsilon-\omega) \
[G_m^+(\epsilon)\gamma^{+-}({\bf q},\omega)- G_m^-(\epsilon)\gamma^{--}({\bf q},\omega)]
\right\},
\end{eqnarray}
where $n_F(\epsilon)=1/\{1+\exp[(\epsilon-\mu)T]\}=\{1-{\rm tanh}[(\epsilon-\mu)2 T]\}/2$
is the Fermi distribution function and we have used
the quasiclassical approximation for matrix elements
(\ref{matr-el-x-quasicl}).

We turn now to a brief summary of results for $\Pi({\bf q},\omega)$ in
various relevant domains of momenta and frequency. Some of these results can be found in
Ref.~\onlinecite{Khaetskii}; we reproduce them here for the sake of completeness.
The polarization operator in the diffusive regime of momenta (and at $T\ll \Delta$)
was already given in
Eq.\ (\ref{diffusive-polarization}).
In the ballistic regime $qR_c\gg 1$,
the expression (\ref{pi-qw}) can be simplified by neglecting
the scalar vertex corrections,
\begin{eqnarray}
\label{pi-qw-ballistic-q}
\Pi({\bf q},\omega)&=&{1\over \pi \ell^2}\sum_{n,m}J^2_{n-m}(q R_c)
\int_{-\infty}^\infty {d\epsilon\over \pi}\ n_F(\epsilon)\
[G_n^+(\epsilon+\omega)+G_n^-(\epsilon-\omega)]\ {\rm Im}\ G^+_m(\epsilon),
\end{eqnarray}
For low temperature and frequency, $\omega,T\ll\Delta$,
the real part of the polarization operator (\ref{pi-qw-ballistic-q}) takes the form
\begin{eqnarray}
{\rm Re}\:\Pi(q\gg R_c^{-1},\omega\to 0)
&=&2\nu_0+
2 \nu_0{8 \omega_c \over 3 \pi \Delta }\, J_0^2(q R_c)\
\left[1-\frac{(\mu-E_N)^2}{\Delta^2} \right]^{3/2}
\label{ballistic-realpolarization}
\end{eqnarray}
Here, the first term~\cite{Aleiner}
arises from Landau levels with $n\neq m$, while
the second term represents the contribution of the $N$th LL ($n=m=N$).
The intra-LL (second) term contains an additional
energy factor ${4\over 3}[1-(\mu-E_N)^2/\Delta^2]$ compared to the
case of diffusive momenta, which is due to the suppression of
vertex corrections at high momenta.
The imaginary part of the polarization operator for $\omega,T\ll\Delta$
has the form
\begin{equation}
\label{ballistic-imaginarypolarization}
   {\rm Im}\ \Pi({\bf q},\omega) = 2 \nu_0 {4\omega \omega_c \over\pi \Delta^2}J_0^2(qR_c)
     \left[1-{(\mu-E_N)^2\over \Delta^2}\right].
\end{equation}
A comparison with Eq.\ (\ref{ballistic-realpolarization}) shows that
${\rm Im}\Pi\ll{\rm Re}\Pi$ in this regime.

It follows from Eq.\ (\ref{ballistic-realpolarization}) that there is an additional
wavevector scale $qR_c\sim \omega_c/\Delta$ in the ballistic regime, where the behavior of
${\rm Re}\Pi$ changes. Specifically, for $q\ll \omega_c/\Delta R_c$
the polarization operator (and hence screening)
is due to the contribution of the $N$-th Landau level,
while at larger $q$ it is due to Landau levels with $n\neq N$.
Only in the latter case, we recover
\begin{equation}
\Pi({\bf q},\omega)\simeq 2\nu_0,
\label{nLL-Pi}
\end{equation}
and thus the standard $B=0$ form of screening.
When the temperature is large compared to the Landau level broadening, $\Delta \ll T \ll \omega_c,$
Eqs.~(\ref{diffusive-polarization}),
(\ref{ballistic-imaginarypolarization}), and the second term of
(\ref{ballistic-realpolarization})
are effectively multiplied by factors $\sim \Delta/T$ due to thermal
averaging. In this case, the real part of
$\Pi({\bf q},\omega)$ takes its zero-$B$ form
under the weaker condition $qR_c\gg \omega_c/T$. This follows from the expression
\begin{eqnarray}
{\rm Re}\:\Pi({\bf q},\omega)
=2\nu_0+
{2\nu_0}{2\omega_c \over \pi T}\ J_0^2(q R_c)\ {\cal Q}\left({\omega \over 2\Delta}\right)\
{\rm cosh}^{-2}\left({E_N-\mu\over 2 T}\right).
\label{ballistic-realpolarization-higherT}
\end{eqnarray}
Here we defined the function
\begin{eqnarray}
{\cal Q}(x)&=&
\int_{-1}^1 dz\ z\ (1-z^2)^{1/2}\left\{z -{{\rm sgn}(z+2x)\over 2}{\rm Re}\,[(z+2x)^2-1]^{1/2}
-{{\rm sgn}(z-2x)\over 2}{\rm Re}\,[(z-2x)^2-1]^{1/2} \right\}.
\label{functionQ(x)}
\end{eqnarray}
The imaginary part for $T\ll \omega_c$ reads
\begin{equation}
\label{Impi-ballistic-q-T<omegac}
{\rm Im}\ \Pi({\bf q},\omega)=
2\nu_0 {2 \omega_c \over \pi T} J^2_0(q R_c)\
{\omega \over 2\Delta} {\cal H}\left({\omega\over 2\Delta}\right){\rm cosh}^{-2}\left({E_N-\mu\over 2 T}\right),
\end{equation}
where ${\cal H}(x)$ is a dimensionless function representing the overlap
of two Landau bands,
\begin{equation}
{\cal H}(x)\equiv \int_{-\infty}^{\infty}{d z}
\left\{{\rm Re}\left[1-(z+x)^2\right]^{1/2}\right\}
\left\{{\rm Re}\left[1-(z-x)^2\right]^{1/2}\right\}.
\label{calH}
\end{equation}

Finally, in the high-T limit, $T\gg \omega_c,$ the imaginary part
of $\Pi({\bf q},\omega)$ becomes independent of $E_N-\mu,$
because of thermal averaging,
\begin{eqnarray}
\label{Impi-ballistic-q}
{\rm Im}\ \Pi({\bf q},\omega)
&\simeq& 2\nu_0{2\omega_c\over \pi \Delta } \sum_{n,m}
\left[{\rm tanh}{E_n+\omega-\mu\over 2 T}- {\rm tanh}{E_n-\mu\over 2 T}\right]\
J^2_{n-m}(q R_c)\  {\cal H}\left({E_n-E_m+\omega\over 2\Delta}\right)
\nonumber \\
&\simeq& 2\nu_0{4 \omega \over \pi \Delta} \sum_k J^2_{k}(q R_c)\
{\cal H}\left({\omega-k\omega_c\over 2\Delta}\right).
\end{eqnarray}
Since ${\cal H}(|x|>1)=0,$
the imaginary part of $\Pi({\bf q},\omega)$ as a function of $\omega$
at $T\gg \omega_c$ consists of a series of peaks (broadened by $\Delta$)
around multiples of the cyclotron frequency.

Importantly, the imaginary part
of the polarization operator is suppressed
at high frequencies, $\omega \gg q v_F$. This follows from
Eq.~(\ref{Impi-ballistic-q}), since $J^2_n(q R_c)$ is exponentially small when $n\gg q R_c$.
This is analogous to the zero-$B$ case, where
\begin{equation}
{\rm Im}\ \Pi({\bf q},\omega;B=0)=2\nu_0 {\omega \over q v_F}\theta(qv_F-\omega)
\label{ImPiB=0}
\end{equation}
with $\theta(x)$ the step function, and
can be traced back to the fact that at high frequencies the magnetic
field becomes almost irrelevant, so that the polarization operator
approaches its zero-$B$ form.~\cite{Aleiner}

\section{Drag resistivity}
\label{dragconductivity}

In a strong magnetic field, $\omega_c\tau_0\gg 1$, the intralayer Hall
resistivity $\rho_{xy}$ dominates over the longitudinal resistivity
$\rho_{xx}$. Therefore, the  drag resistivity is given by
\begin{equation}
\rho_{xx}^D\simeq
\rho_{xy}^{(1)}\ \sigma^D_{yy}\ \rho_{yx}^{(2)}.
\label{rhoxx-drag}
\end{equation}
Using Eq.~(\ref{drag-conductivity-retarded}), we get the expression for the
longitudinal component of the drag resistivity in a strong magnetic field,
\begin{eqnarray}
\rho_{xx}^D
&=&-\frac{B}{e n_{e1}}\frac{B}{e n_{e2}}{1 \over 8\pi}\int_{-\infty}^\infty
{d\omega \over 2 T\sinh^2(\omega/2 T)}
\nonumber \\
&\times&\int {d^2{\bf q}\over (2 \pi)^2}
\Gamma_y^{(1)}({\bf q},\omega,B)\Gamma_y^{(2)}({\bf
  q},\omega,-B)|U_{12}({\bf q},\omega)|^2.
\label{rho-general-B}
\end{eqnarray}
The overall minus sign in  Eq.\ (\ref{rho-general-B}) is due to the
relation $\rho_{xy}=-\rho_{yx}$. It follows that
for identical layers, the longitudinal
(${\bf \Gamma}\propto \hat{\bf q}$) component $\Gamma_{||}$ of the triangle vertex
gives rise to {\it negative} drag, since $\Gamma_{||}(-B)=\Gamma_{||}(-B),$
while the transverse (${\bf \Gamma}\propto \hat{\bf z} \times \hat{\bf q}$)
component $\Gamma_{\perp}$ yields {\it positive} drag,
$\Gamma_{\perp}(-B)=-\Gamma_{\perp}(-B).$

Since the upper limit of the momentum integration in
(\ref{rho-general-B}) is effectively set by the inverse interlayer
distance, $a^{-1}$, the behavior of the transresistivity will
essentially dependent on the relation between $R_c$ and
$a$. Below, we mainly concentrate on the {\it ballistic} regime
\begin{equation}
  \omega_c/\Delta\ll R_c/a\ll N\Delta/\omega_c,
  \label{ballistic-condition}
\end{equation}
which we consider as
most relevant experimentally. In Sec.~\ref{regimes} we will briefly
consider other situations and discuss the evolution of the
transresistivity with decreasing interlayer distance, from the
diffusive ($R_c/a\ll 1$) to the ultra-ballistic ($R_c/a\gg
N\Delta/\omega_c$) regime.

\subsection{Ballistic regime: Low temperatures ($T\ll \Delta$)}
\label{calculation}

In the low-temperature limit, the expressions derived for the triangle
vertex ${\bf \Gamma}({\bf q},\omega)$ at $\omega\ll \Delta$
are sufficient, because frequencies in Eq.~(\ref{rho-general-B}) are
restricted to $\omega\lesssim T \ll \Delta$.
Let us analyze which of the contributions to the triangle vertex dominates,
depending on the relation between $q$ and $1/R_c$.

In the diffusive range of momenta, $qR_c\ll 1$, the leading
contribution to the triangle vertex is given by
Eq.~(\ref{diffusive-omega}); its magnitude can be estimated as
\begin{eqnarray}
\Gamma&\sim & {\omega k_F \over \Delta^2 q R_c}.
 \label{Gamma1-diff}
\end{eqnarray}

In the ballistic regime, $qR_c\gg 1,$ we have three competing
contributions (see Sec.~\ref{ballistic}),
\begin{eqnarray}
\Gamma^{(1/qR_c)} &\sim & {\omega k_F \over \Delta^2 (q R_c)^2},
 \label{Gamma1-ball}\\
\Gamma^{(\Delta/\omega_c)}&\sim & {\omega k_F \over \Delta \omega_c q R_c}
\sim q R_c {\Delta\over \omega_c}\Gamma^{(1/qR_c)},
\label{Gamma2-ball}\\
\Gamma^{(q/k_F)}&\sim & {\omega \over \Delta^2 R_c}
\sim {(q R_c)^2\over N}\Gamma^{(1/qR_c)}
\sim {q \over k_F} {\omega_c\over\Delta}\Gamma^{(\Delta/\omega_c)}.
\label{Gamma3-ball}
\end{eqnarray}
Comparing these expressions, we find that the first contribution,
$\Gamma^{(1/qR_c)}$, dominates for $qR_c\ll \omega_c/\Delta$,
the second one, $\Gamma^{(\Delta/\omega_c)}$, is dominant
for $\omega_c/\Delta \ll q R_c \ll N \Delta/\omega_c$, while the last
contribution,  $\Gamma^{(q/k_F)}$,
becomes the largest one for $ q R_c \gg N \Delta/\omega_c$.
This is valid provided that the Landau level index $N$ is sufficiently
large, $N>(\omega_c/\Delta)^2$.
We will assume below that this condition
is fulfilled.

Splitting the momentum integral in Eq.\ (\ref{rho-general-B}) into
three parts, corresponding to regions of different behavior of the
triangle vertex and polarization
operator, we present the transresistivity in the following form,
\begin{eqnarray}
\rho_{xx}^D&\simeq& {e^2B^2\over 8\pi^3}
\left(k_F\over \kappa_0^2 n_{e}\Delta^2\right)_1
\left(k_F\over \kappa_0^2 n_{e} \Delta^2\right)_2
\int_{-\infty}^\infty d\omega {\omega^2 \over 2T \sinh^2(\omega/2 T)}
I(\omega), \label{rho-w-integral}\\[0.3cm]
I(\omega)&=& I_{\rm I}(\omega)+I_{\rm II}(\omega)+I_{\rm III}(\omega),
\label{Iwsplit}
\end{eqnarray}
where the subscript $l=1,2\ $ in $\ \left(\ldots\right)_l$ refers to the
layer $l$,
and the contributions $I_{\rm I}$, $I_{\rm II}$, and $I_{\rm III}$ in
(\ref{rho-w-integral}) are determined by the momentum domains $qR_c\ll
1$, $1\ll qR_c\ll \omega_c/\Delta$, and $qR_c\gg \omega_c/\Delta$,
respectively. The corresponding expressions are given in
Appendix~\ref{app:drag-split}. Estimating all three terms, we find
[see Eq.~(\ref{sum3I})] that the
leading contribution is given by the last term,
\begin{eqnarray}
I(\omega)&\simeq& I_{\rm III}(\omega)=
{1\over 2 \pi^3 a^2 R_c^2} \ln\left({R_c \Delta \over a \omega_c}\right)
\left\{\frac{16}{\omega_c\Delta^2} (\mu-E_N)[\Delta^2-(\mu-E_N)^2]\right\}_1
\left\{\frac{16}{\omega_c\Delta^2} (\mu-E_N)[\Delta^2-(\mu-E_N)^2]\right\}_2,
\nonumber \\
&\sim&
{1\over a^2 R_c^2} \left({\Delta\over \omega_c}\right)^2
\ln\left(\frac{R_c \Delta}{a \omega_c}\right).
\end{eqnarray}
Therefore, for $T\ll \Delta$ we get for identical layers
\begin{eqnarray}
\rho_{xx}^D
&=& {32\over 3\pi^2 e^2}{1\over (k_F a)^2 (\kappa_0 R_c)^2}
\left({T\over \Delta}\right)^2\ln\left(\frac{R_c \Delta}{a
    \omega_c}\right)
\left(\frac{\mu-E_N}{\Delta}\right)^2 \left[1-{(\mu-E_N)^2\over
    \Delta^2}\right]^{2}.
\label{rho-leading}
\end{eqnarray}
Thus at low temperature $T\ll \Delta$,
the drag resistivity scales with the magnetic field and temperature as
\begin{equation}
\rho_{xx}^D\propto T^2 B \ln (B_*/B),
\label{scaling-low-T}
\end{equation}
where $B_*\sim (mc/e)(v_F^2/a^2\tau_0)^{1/3}$ sets the upper boundary for
the considered ballistic regime on the magnetic field axis.

If $R_c$ differs slightly between the two layers (i.e., the concentrations
are slightly different) so that $\delta R_c/a \ll 1$, the above calculation
fully applies, with the only change in Eq.~(\ref{rho-leading})
\begin{eqnarray}
\left(\frac{\mu-E_N}{\Delta^2}\right)^2 \left[1-{(\mu-E_N)^2\over
    \Delta^2}\right]^{2}
&\to& \left(\frac{\mu-E_N}{\Delta} \left[1-{(\mu-E_N)^2\over
    \Delta^2}\right]\right)_1\\
&\times& \left(\frac{\mu-E_N}{\Delta}
\left[1-{(\mu-E_N)^2\over \Delta^2}\right]\right)_2.
\label{change1}
\end{eqnarray}
This yields an {\it oscillatory sign} of the drag.
For identical layers the drag is {\it positive}, at variance with
Ref.~\onlinecite{Oppen}.
This is because the leading term here originates from the component
$\Gamma_{\perp}$ of the triangle vertex transverse to the wave vector
${\bf q}$ (i.e. directed along $\hat{{\bf z}}\times\hat{\bf q}$). For a more
detailed discussion of the sign of drag in different regimes, see
Sec.~\ref{regimes}.

If $a\ll \delta R_c \ll R_c \Delta/\omega_c$,
the calculation still applies, but the argument of the logarithm
changes,
\begin{equation}
\ln\left(\frac{R_c}{a}\frac{\Delta}{\omega_c}\right)\to
\ln\left(\frac{R_c}{\delta R_c} \frac{\Delta}{\omega_c}\right).
\label{change2}
\end{equation}

\subsection{Ballistic regime: Arbitrary $T/\Delta$}
\label{highTdrag}

\begin{figure}
\centering
\epsfxsize7cm\epsfbox{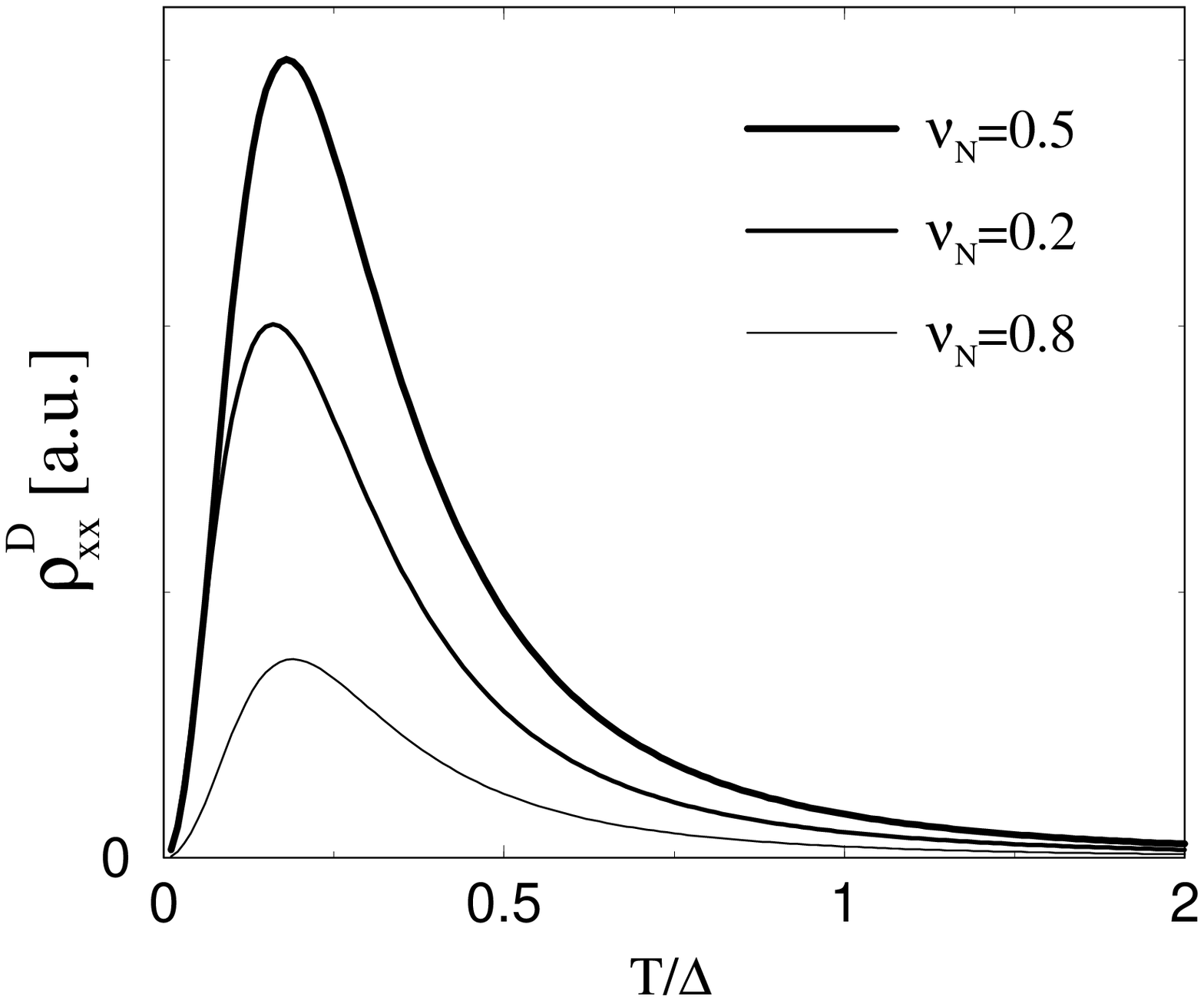}\hskip0.5cm
\epsfxsize7.4cm\epsfbox{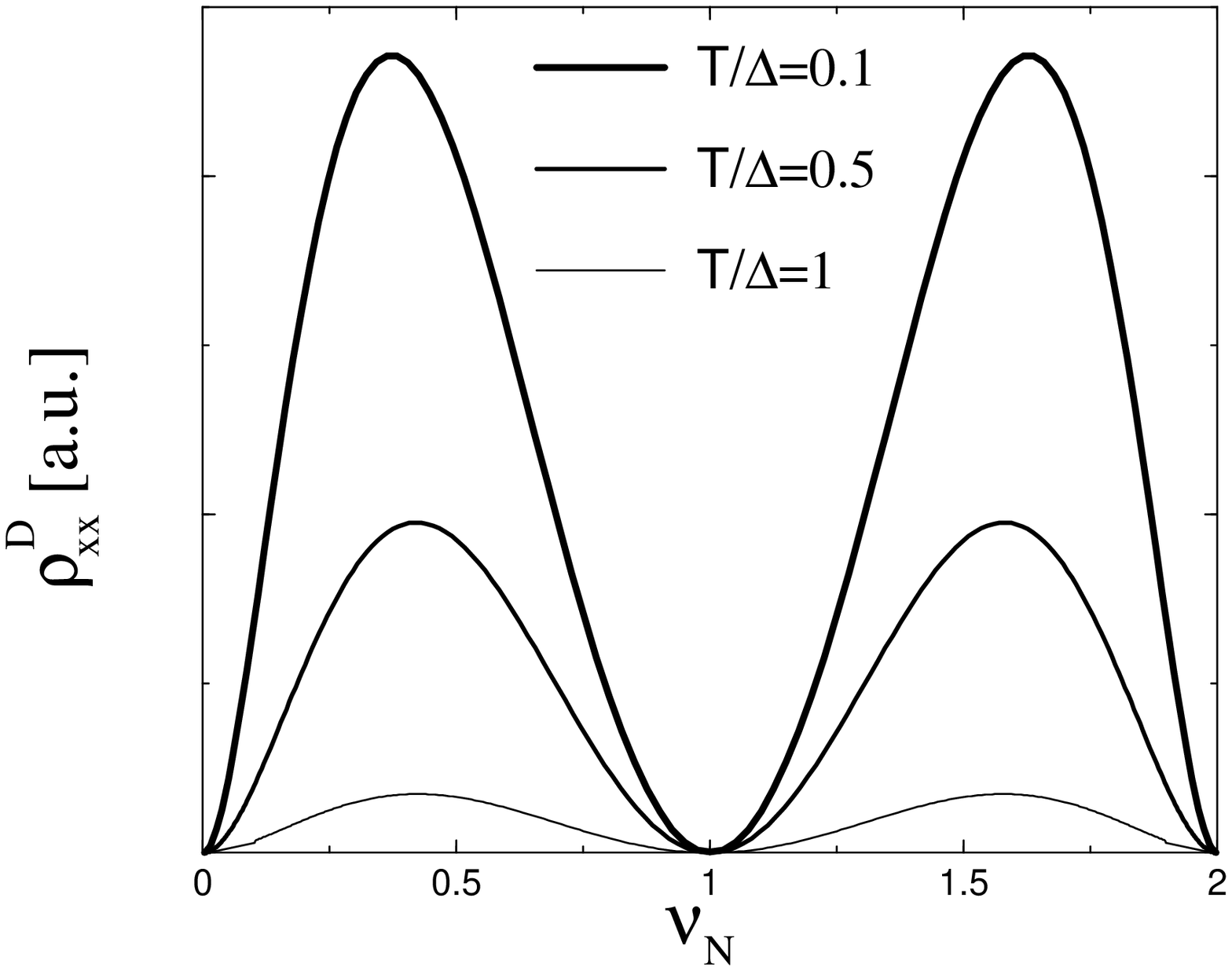}\vskip .3truecm
\caption{ Low-temperature drag for identical layers.
Left panel: temperature dependences of the ${\cal O}(\Delta/\omega_c)$-term
in $\rho_{xx}^D(T)$ for different values of the filling factor of the highest LL,
$\nu_N=0.5,0.2, 0.8$ (from top to bottom). Right panel:
dependence of the ${\cal O}(\Delta/\omega_c)$-term on the filling factor $\nu_N$
 for different values of temperature,
$T/\Delta=0.1,\ 0.5,\ 1$ (from top to bottom).}
\label{fig-num1}
\end{figure}

Having identified the leading contribution (coming from $\Delta/\omega_c$-term)
to drag for
temperatures small compared to the LL width $\Delta$, we generalize
the obtained result
to the case of larger $T$ (and correspondingly $\omega$).
As discussed in Appendix\ \ref{app:drag-split}, the only difference
in the momentum integral in the $\Delta/\omega_c$-term is the replacement
$\Delta \to T$ under the argument of logarithm.
Using Eq.~(\ref{finite-w-2}) and assuming that the difference
in $R_c$ between the two layers is not too large, $\delta R_c\ll a$,
we express the ${\cal O}(\Delta/\omega_c)$-contribution to the
transresistivity as
\begin{eqnarray}
\left(\rho_{xx}^D\right)^{(\Delta/\omega_c)}
&=& {4\over \pi^4 e^2}{1\over (k_F a)^2 (\kappa_0 R_c)^2}
\ln\left(\frac{R_c {\rm max}[\Delta,T]}{a \omega_c}\right)\nonumber \\
&\times&\int_{-\infty}^{\infty}\frac{d \omega}{2T \sinh^2(\omega/2T)}
[{\tilde {\cal F}}(\omega,\mu,T)]_1
[{\tilde {\cal F}}(\omega,\mu,T)]_2,
\label{pho-leading-full}
\end{eqnarray}
where ${\tilde {\cal F}}(\omega,\mu,T)$ is a dimensionless function of
$\omega/\Delta,$ $(\mu-E_N)/\Delta,$ and $T/\Delta,$
\begin{eqnarray}
{\tilde {\cal F}}(\omega,\mu,T)&\equiv & \int_{-\infty}^{\infty}{d\epsilon\over \Delta}
\left(\tanh\frac{\epsilon+\omega/2-\mu}{2T}
- \tanh\frac{\epsilon-\omega/2-\mu}{2T}
\right)
\nonumber \\
&\times &
\frac{\epsilon-E_N}{\Delta}\left[1-{(\epsilon+\omega/2-E_N)^2 \over
    \Delta^2}\right]^{1/2}
\left[1-{(\epsilon-\omega/2-E_N)^2\over \Delta^2}\right]^{1/2}.
\label{Fwm}
\end{eqnarray}
For arbitrary $T/\Delta\sim 1$ this can only be calculated numerically.
In Fig.~\ref{fig-num1} we present the results for the temperature dependence
of the ${\cal O}(\Delta/\omega_c)$-contribution to drag as well as for its
dependence on the filling factor $\nu_N$ of the highest (partially filled, $0<\nu_N<2$) LL.
It is worth mentioning that when temperature is varied at fixed filling factor
(as in typical experiments), the chemical potential is varying as well, $\mu=\mu(\nu_N,T),$
which is taken into account in Fig.~\ref{fig-num1}.

Consider the regime of temperatures large compared to the LL
width, $T\gg \Delta$.
In this situation the LLs will be broadened by the temperature, so that
typically $\mu-E_N$ will be of order $T$  and thus much larger than
$\Delta$.
Expanding the $\tanh$-terms in (\ref{Fwm}) in $\omega<2\Delta\ll T$
and $\epsilon-E_N<\Delta\ll T$,
we arrive at
\begin{eqnarray}
\left(\rho_{xx}^D\right)^{(\Delta/\omega_c)}
&\simeq& {4\over \pi^4 e^2}{1\over (k_F a)^2 (\kappa_0 R_c)^2}
\ln\left(\frac{R_c T}{a \omega_c}\right)\nonumber \\
&\times&
\left(
\sinh{E_N-\mu\over 2T} \cosh^{-3}{E_N-\mu\over 2T}
\right)_1
\left(
\sinh{E_N-\mu\over 2T} \cosh^{-3}{E_N-\mu\over 2T}
\right)_2
\nonumber \\
&\times&
\left({\Delta\over T}\right)^3
\int_{-\infty}^{\infty}\frac{d \omega}{2\Delta} [{{\cal F}}(\omega/2\Delta)]_1
[{{\cal F}}(\omega/2\Delta)]_2,
\label{rho-leading-high-T}
\end{eqnarray}
where ${\cal F}(x)$ is a dimensionless function similar to Eq.~(\ref{calH}).
It also describes the overlap of two shifted Landau levels,
but has an additional factor $[(\epsilon-E_N)/\Delta]^2$
arising from the particle--hole asymmetry due to LL quantization,
\begin{equation}
{{\cal F}}(x)\equiv \int_{-\infty}^{\infty}
{d z}\  z^2
\left\{{\rm Re}\left[1-(z+x)^2\right]^{1/2}\right\}
\left\{{\rm Re}\left[1-(z-x)^2\right]^{1/2}\right\}
\label{tildeFw}
\end{equation}
The contribution (\ref{rho-leading-high-T}) scales as
\begin{equation}
\rho_{xx}^D\propto T^{-3}B^{7/2}\ln (T/B^2).
\label{high-T-scaling}
\end{equation}

Since the above ${\cal O}(\Delta/\omega_c)$-term falls off quickly at
$T\gg \Delta,$ we should analyze the contributions of the other terms.
Let us first calculate the ``conventional'' term ${\cal O}(q/k_F),$
substituting Eq.~(\ref{gamma-qkf}) in Eq.~(\ref{rho-general-B}).
Remarkably, the \textit{strong}-$B$ expression for the $q/k_F$-contribution to
drag \textit{resistivity} reduces to the standard
\textit{zero}-$B$ form~\cite{MacDonald,Kamenev,Flensberg95},
\begin{eqnarray}
\left(\rho_{xx}^D\right)^{(q/k_F)}
&=&\frac{1}{4\pi e^2 n_{e1}n_{e2}}
\int_{-\infty}^\infty {d\omega \over 2 T\sinh^2(\omega/2 T)}
\nonumber \\
&\times&
\int {d^2{\bf q}\over (2 \pi)^2} q^2
\left[{\rm Im}\ \Pi({\bf q},\omega)\right]_1 \
\left[{\rm Im}\ \Pi({\bf q},\omega)\right]_2 \
|U_{12}({\bf q},\omega)|^2.
\label{rho-q/kf-standard}
\end{eqnarray}
Here all the information about the magnetic field is encoded
in $\Pi({\bf q},\omega),$ Eq.~(\ref{pi-qw}).

For $T\gg \Delta,$ expanding the $\tanh$-terms in ${\rm Im}\ \Pi({\bf q},\omega)$
just as before we find
\begin{eqnarray}
\left(\rho_{xx}^D\right)^{(q/k_F)}
&\simeq& {3\zeta(3)\over 2 \pi^4 e^2}{1\over (k_F a)^4(\kappa_0 R_c)^2 }
\left({\omega_c\over \Delta}\right)^2\ {\Delta \over T}\nonumber \\
&\times&\left(\cosh^{-2}{E_N-\mu\over 2T}\right)_1
\left(\cosh^{-2}{E_N-\mu\over 2T}\right)_2
\int_{-\infty}^{\infty}\frac{d \omega}{2\Delta} [{{\cal H}}(\omega/2\Delta)]_1
[{{\cal H}}(\omega/2\Delta)]_2,
\label{rho-very-high-T}
\end{eqnarray}
where $\zeta(x)$ is the Riemann zeta-function [$\zeta(3)\simeq 1.202$] and ${\cal H}(x)$ is
defined in Eq.~(\ref{calH}).
This contribution scales as
\begin{equation}
\rho_{xx}^D\propto T^{-1}B^{7/2}.
\label{very-high-T-scaling}
\end{equation}

The slower fall-off of the ${\cal O}(q/k_F)$-contribution
(\ref{rho-very-high-T}) as compared to
(\ref{rho-leading-high-T}) can be traced back to the different nature of
the particle--hole asymmetry underlying these two contributions.
Specifically, the ${\cal O}(\Delta/\omega_c)$-term
(\ref{rho-very-high-T}) is governed
by the particle--hole asymmetry due to the LL quantization.
This is reflected by
the factor $\epsilon-E_N$ in (\ref{Fwm}) which after thermal averaging,
translates into a factor in Eq.~(\ref{rho-leading-high-T}) which is asymmetric
in $E_N-\mu$.
On the other hand, the ``conventional'' $q/k_F$ contribution is due to the
curvature of zero-$B$ spectrum and therefore is symmetric in $\epsilon-E_N$
(and in $E_N-\mu$ after thermal averaging).
In both cases the fall-off of drag at $T\gg \Delta$ is due to the
absence of electronic states outside the Landau band
(for $|\epsilon-E_N|>\Delta$). However, the thermal averaging
of the odd function of $\epsilon-E_N$ yields an additional
factor $\Delta/T$ for each $\Delta/\omega_c$-triangle vertex, at
variance with the case of an even function of $\epsilon-E_N$ determining
${\cal O}(q/k_F)$-contribution.

Finally, we evaluate the contribution of ${\cal O}(1/qR_c)$-term.
On the one hand, the thermal averaging suppresses each ${\bf\Gamma}^{(1/qR_c)}$
vertex by the factor $(\Delta/T)^2$, similarly to the
${\cal O}(\Delta/\omega_c)$-term. This is again because of the
particle--hole asymmetry due to the LL quantization.
On the other hand, the peculiarity of the finite-$T$ screening
gives rise to a factor $(T/\Delta)^2$ in the momentum integral involving
the ${\cal O}(1/qR_c)$-term, see Appendix~\ref{app:drag-split}.
The remaining frequency integral yields the factor $T/\Delta$,
since the allowed frequencies are restricted by $|\omega|<2\Delta\ll T$.
As a result, the contribution of this term to the
drag resistivity is inversely proportional to temperature for $T\gg\Delta,$
similarly to the conventional $q/k_F$-contribution.
For simplicity we restrict ourselves to the case of identical layers,
where we get
\begin{eqnarray}
\left(\rho_{xx}^D\right)^{(1/qR_c)}
&\simeq& -{c^{(1/qR_c)}\over e^2}{1\over (k_F a)^2(\kappa_0 R_c)^2} {\Delta\over T}
\sinh^2{E_N-\mu\over 2T} \cosh^{-2}{E_N-\mu\over 2T},
\label{rho-more-leading-high-T}
\end{eqnarray}
where $c^{(1/qR_c)}$ is a constant of order unity,
\begin{eqnarray}
c^{(1/qR_c)}&=&
{4\over \pi} \int_{-\infty}^{\infty} d x \left[{\cal P}(x)\right]^2 {\cal W}(x),
\label{c1qRc}
\end{eqnarray}
with the function ${\cal W}(x)$ defined in Eq.~(\ref{W(y)}) and
\begin{equation}
{\cal P}(x)=\int_{-\infty}^{\infty}dz\ z\ (z+x)\ {\rm Re}[1-(z+x)^2]^{1/2} \
\left\{{\rm Re}[1-(z-x)^2]^{1/2}\right\}^2.
\label{calP(x)}
\end{equation}
This contribution scales as
\begin{equation}
\rho_{xx}^D\propto -T^{-1}B^{5/2}.
\label{1/qRc-high-T-scaling}
\end{equation}
We thus conclude that the ${\cal O}(1/qR_c)$-contribution wins over the
${\cal O}(\Delta/\omega_c)$-contribution
for $$T>\Delta\ln^{1/2}\left(\frac{R_c \Delta}{a \omega_c}\right)\equiv T_*.$$
Comparing (\ref{rho-very-high-T}) and (\ref{rho-more-leading-high-T}),
we have
\begin{equation}
\frac{{\cal O}(q/k_F)}{|{\cal O}(1/qR_c)|}\sim
{1\over (k_F a)^2}(\omega_c/\Delta)^2
\sim \left({R_c\over a} {\omega_c\over N\Delta} \right)^2\ll 1,
\end{equation}
as follows from Eq.~(\ref{ballistic-condition}).
Therefore the ${\cal O}(1/qR_c)$-contribution dominates the drag resistivity
in the intermediate range of temperature. This contribution oscillates with
changing the filling factor of the two layers;
however, it is negative for matching densities,
unlike the ${\cal O}(\Delta/\omega_c)$-contribution.

For higher temperatures, $T>\omega_c,$
the terms related to the LL particle-hole asymmetry fall off rapidly
due to the thermal averaging involving many LLs and
thus the $q/k_F$-term
(i.e. the ``conventional'' contribution to the drag resistivity)
soon becomes dominant. The drag resistivity is then always positive, independently of
the difference in filling factors of the two layers.
It monotonously increases
with increasing $T$ and takes the form
\begin{eqnarray}
\rho_{xx}^D
&\simeq& {8 \zeta(3)\over \pi^2 e^2}{1\over (k_F a)^4 (\kappa_0 R_c)^2}
{\omega_c\over \Delta} \left({T\over \omega_c}\right)^2\
\int_{-\infty}^{\infty}\frac{d \omega}{2\Delta} [{{\cal H}}(\omega/2\Delta)]_1
[{{\cal H}}(\omega/2\Delta)]_2\nonumber \\
&\sim&
 {1 \over e^2(k_F a)^{2}(\kappa_0 a)^{2}}\left({T \over E_F}\right)^2
{\omega_c \over \Delta}
\propto T^2B^{1/2}.
\label{T>w}
\end{eqnarray}
This is almost the same result that is found in zero magnetic field
\cite{MacDonald,Kamenev,Flensberg95}; the only difference is an extra factor $\sim
\omega_c/\Delta \propto B^{1/2}$. The reason for the emergence of
the zero-$B$ result is physically very transparent. Characteristic
frequencies $\omega\sim T\gg \omega_c$ set a characteristic time scale
$T^{-1}$, which is much smaller than the time of the cyclotron
revolution. At such times the electron motion is essentially
unaffected by the magnetic field. The magnetic field enters, however,
through the density of states $\nu$ inside the LL, which determines the
characteristic magnitude of ${\rm Im}\Pi$, and thus of ${\bf \Gamma}$, see
Eq.~(\ref{gamma-qkf}). The $\omega$-integration in
(\ref{rho-general-B}) thus results in an effective averaging of
$\nu^2$, yielding the factor $\omega_c/\Delta$.
It is worth mentioning that, for the same reason,
the longitudinal resistivity $\rho_{xx}$ of a single layer is also
enhanced by such a factor in the regime $T\gg\omega_c\gg\Delta$
as compared to its zero-$B$ value, see e.g. Ref.~\onlinecite{Vavilov}.

For still higher temperature, $T\gg v_F/a,$ the
quadratic-in-$T$ dependence of the drag resistivity crosses over into the
linear-in-$T$ drag. This occurs because of the suppression of the imaginary
part of the polarization operator [determining the $q/k_F$-triangle, Eq.~(\ref{gamma-qkf})]
at $\omega\gg qv_F$, see Eq.~(\ref{ImPiB=0}). As a result, the domain of $\omega$-integration
is effectively restricted by $\omega\alt v_F/a$ (since $q\alt 1/a$), yielding
the replacement $T^2 \to T v_F/a$ as compared to the case of $\omega_c\ll T\ll v_F/a$,
$$\rho_{xx}^D\propto T B^{1/2}.$$

Before closing this subsection, it is worth mentioning that
in the above consideration we have neglected the contribution of magnetoplasmons
to the drag (see Ref.~\onlinecite{Khaetskii} for details).
While this contribution may become important for very high temperatures, $T\gg \omega_c,$
it is negligibly small at relatively low $T\sim \Delta,$ which is the range
of our main interest in the present paper.

\subsection{Comparison with Experiment}
\label{experiment}

\begin{figure}
\centering
\epsfxsize8cm\epsfbox{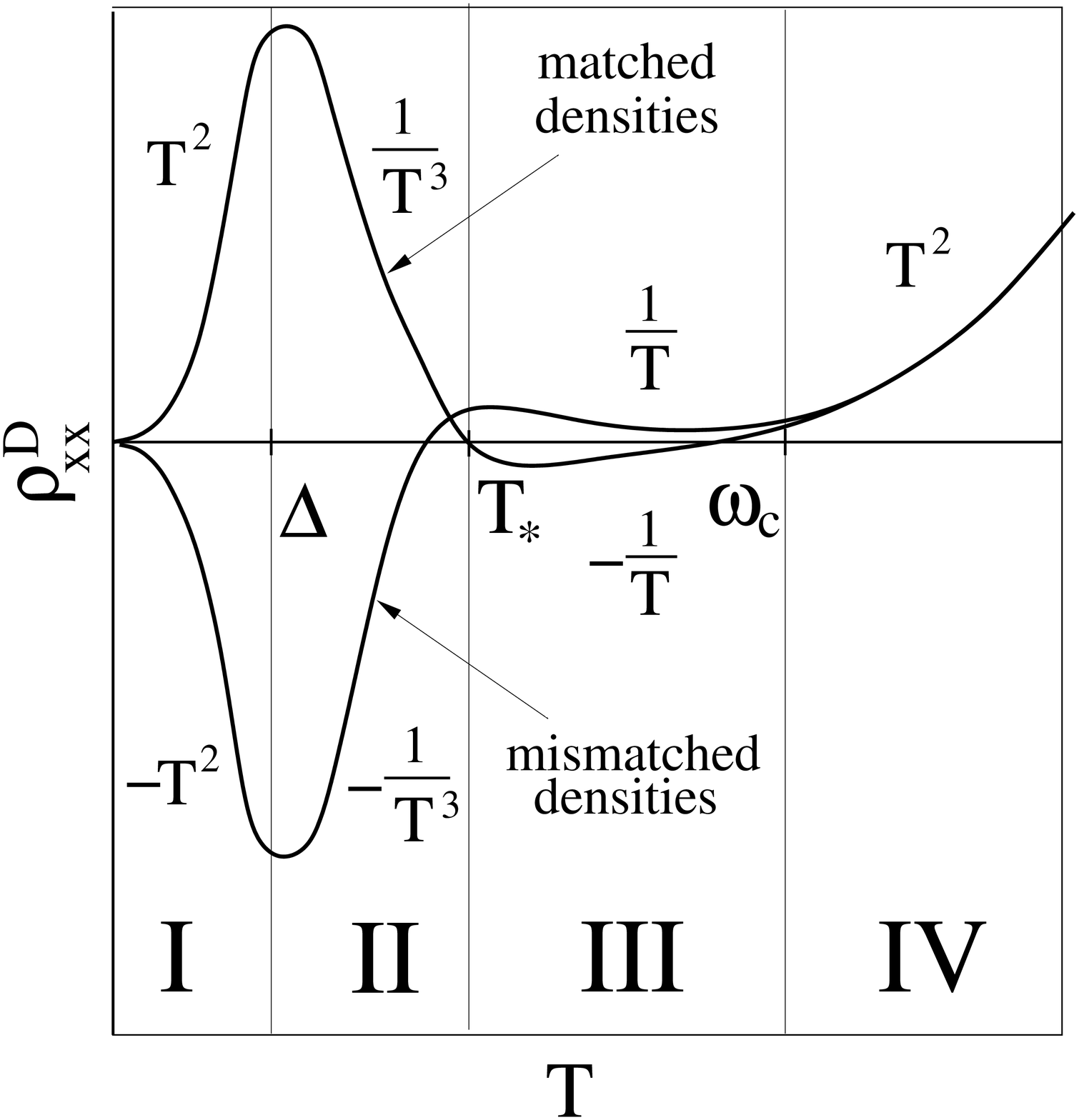} \vskip .3truecm
\caption{Schematic temperature dependence of drag in the ballistic
  regime for matched and mismatched densities. In the latter case the
  mismatch is chosen such that the drag is negative at low $T$ (see
  text). Scaling of $\rho_{xx}^D$ with temperature in different
  regions is indicated: I --  Eq.~(\ref{scaling-low-T}), II -- Eq.~(\ref{high-T-scaling}),
III -- Eq.~(\ref{1/qRc-high-T-scaling}), and IV -- Eq.~(\ref{T>w}). }
\label{fig6}
\end{figure}

In this subsection we compare the results for the drag in the
ballistic regime obtained above with experimental findings. We have
found a sequence of different regimes of the
temperature behavior of $\rho_{xx}^D$,  see
Eqs.~(\ref{scaling-low-T}), (\ref{high-T-scaling}),
(\ref{1/qRc-high-T-scaling}), (\ref{T>w}). All these results are
schematically summarized in Fig.~\ref{fig6}. The upper curves there
depicts $\rho_{xx}^D(T)$ for equal densities, whereas the lower curve
corresponds to a mismatch in densities chosen in such a way that the
Fermi energy is located in the upper half of the Landau
band in one layer, and in the lower half in the other layer.
As has been already emphasized, the drag at low temperatures is
positive for matched and negative for mismatched densities.
This sign of the oscillatory drag can be traced back to the fact that the
dominant contribution to the triangle vertex is given by ${\bf
  \Gamma}^{(\Delta/\omega_c)}$, which is transverse with respect
to the momentum ${\bf q}$.

We now compare these results with a most recent and detailed study by
Muraki {\it et al.}\cite{Muraki}
of the Coulomb drag in the regime of high Landau
levels. A comparison of our Fig.~\ref{fig6} with Fig.~3 of
Ref.~\onlinecite{Muraki} reveals a remarkable agreement between the
experimental findings and our theoretical results. In both the theory
and the experiment,
(i) $\rho_{xx}^D(T)$ shows a sharp peak at low
temperatures;
(ii) the sign of the drag in this temperature range
oscillates as a function of the filling factor of one layer (at fixed
filling factor of the other layer);
(iii) the low-$T$ drag is positive for equal
filling factors and negative when the Fermi energy in one layer is in
the upper half and in the other layer in the lower half of the
Landau band;
(iv) the high-$T$ drag is always positive, independently of the difference
in filling factors of two layers and increases monotonically with increasing $T$.
Furthermore, it was observed by Muraki {\it et al} (see
Fig.2 of Ref.~\onlinecite{Muraki})
that in the low-temperature regime of initial increase of
$\rho_{xx}^D$, as well as in the high-temperature regime of ``normal''
drag, the drag resistivity can be described by an
empirical scaling law,
\begin{equation}
\rho_{xx}^D\propto \left({n \over B}\right)^{-2.7}f(T/B).
\label{scaling-low-T-exp}
\end{equation}
Our results for the low-temperature, (\ref{scaling-low-T}), and
high-temperature, (\ref{T>w}), increase of $\rho_{xx}^D$ are in a nice
correspondence with this prediction, with $f(x)\sim x^2$.

The magnitude of the low-temperature peak in
the drag resistivity that follows from our theory also
agrees with the experiment.
Specifically, estimating Eq.~(\ref{rho-leading}) at $T=0.25 \Delta$
and $[(\mu-E_N)/\Delta]^2=1/2$ by making use of typical
experimental parameters,
$\kappa_0 \sim k_F\sim 10^8\ $m$^{-1}, \ a\sim 10^{-8}$m, $R_c \sim 10^{-7}$m,
we find $\rho_{xx}^D\sim 1\Omega,$ in good agreement with the result of
Ref.~\onlinecite{Muraki}.

There is however a difference between our result
(\ref{scaling-low-T}) for the low-temperature scaling of drag
and the interpretation of low-$T$ data in Ref.~\onlinecite{Muraki}.
Specifically, while we find $T^2$ scaling in this regime, Muraki
{\it et al}.\ fit the data to an exponential (activation-type)
dependence, arguing that localized states are responsible for the
low-temperature ``anomalous peak'' in $\rho_{xx}^D(T)$.
We do not expect, however, that localization plays an important
role in the regime of high Landau levels at realistic temperatures.
Indeed, as is seen from
Fig.1 of Ref.~\onlinecite{Muraki}, the resistivity for filling factors
$\nu\gtrsim 10$ has a shape as predicted by SCBA, without developed
Hall plateaus. Also, the fit of the low-$T$ behavior of $\rho_{xx}^D$
to the activated over a single decade is not unambiguous; the same
data could be quite well fitted to the $T^2$ power law. In other
words, we believe that our theory based on SCBA and not including
quantum localization effects is sufficient to explain the most salient
experimental observations of Ref.~\onlinecite{Muraki}:
the ``anomalous'' drag with oscillatory sign at low
temperatures and the ``normal'' positive drag at high $T$.

\subsection{Evolution of $\rho_{xx}^D(T)$ with varying interlayer
  distance: From the diffusive to the ultra-ballistic limit.}
\label{regimes}

\begin{figure}
\centering
\epsfxsize17cm\epsfbox{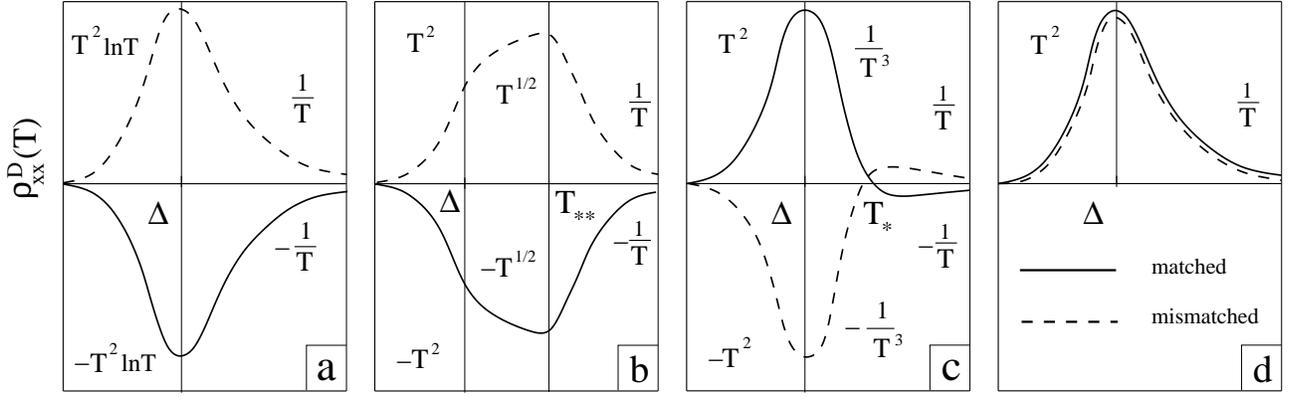} \vskip .3truecm
\caption{Schematic temperature dependence of low-temperature drag in different
regimes:
a) diffusive, $R_c/a\ll 1$;
b) weakly ballistic, $1 \ll R_c/a \ll \omega_c/\Delta$;
c) ballistic, $\omega_c/\Delta \ll R_c/a \ll N\Delta/\omega_c$;
d) ultra-ballistic, $N\Delta/\omega_c\ll R_c/a$.
 }
\label{fig7}
\end{figure}

As discussed in the beginning of Sec.~\ref{calculation},
the form of the drag resistivity $\rho_{xx}^D(T)$ depends on
the value of the ratio $R_c/a.$ In the above we have concentrated
on the regime $\omega_c/\Delta \ll R_c/a \ll N\Delta/\omega_c,$
which can be termed ``ballistic'' and which we believe to be
most relevant to a typical experiment.
In this subsection we briefly describe the results obtained
for other regimes.
Specifically, with increasing $R_c/a$ we identify the following four regimes:
i) diffusive, $R_c/a\ll 1$,
ii) weakly ballistic, $1 \ll R_c/a \ll \omega_c/\Delta$,
iii)   ballistic, $\omega_c/\Delta \ll R_c/a \ll N\Delta/\omega_c$,
and
iv) ultra-ballistic, $N\Delta/\omega_c\ll R_c/a$.
In all regimes, the temperature-dependence of the drag resistivity
is non-monotonous: the absolute value of $\rho_{xx}^D(T)$ shows a peak
around $T\sim \Delta$ and increases again at $T\gg \omega_c.$
However, the $T-$ and $B-$ dependences of $\rho_{xx}^D$, as well
as the sign of the low-temperature peak
(the high-temperature drag is always positive),
are specific for each particular regime, as illustrated in Fig.~\ref{fig7}
and summarized below.

\textit{Diffusive regime}, $R_c/a\ll 1$.
In the diffusive regime, the drag at not too high temperatures, $T\ll \omega_c,$
is governed by the diffusive rectification, Eqs.~(\ref{diffusive-omega})
and (\ref{diffusive-oppen}). As a result, the sign of the drag at $T\sim \Delta$
oscillates but is opposite to what we found above for the
ballistic regime: the drag is negative for equal densities.\cite{Oppen}
At the ``slopes'' of the peak, $\rho_{xx}^D$ scales with $T$ and $B$
in the following way
\begin{equation}
\label{peak-diffusive-scaling}
\rho_{xx}^D \propto \left\{
\begin{array}{ll}
-\ T^2\ \ln (T B^{3/2}), & \qquad \qquad T\ll \Delta, \\[0.2cm]
-\ T^{-1} B^{3/2}\ \ln B, & \qquad \qquad  T\gg \Delta,
\end{array}
\right.
\end{equation}
where the sign corresponds to the case of matching densities.

\textit{Weakly ballistic regime}, $1 \ll R_c/a \ll \omega_c/\Delta$.
This regime is qualitatively similar to the diffusive regime.
The peak at $T\sim \Delta$ is governed now by the ${\cal O}(1/qR_c)$-term
in the triangle vertex,
resulting in
\begin{equation}
\label{peak-weakly-ball-scaling}
\rho_{xx}^D \propto \left\{
\begin{array}{ll}
-\ T^2\ B^{-5/4}, & \qquad \qquad T\ll \Delta, \\[0.2cm]
-\ T^{1/2}\ B^{-1/2}, & \qquad \qquad \Delta \ll T \ll T_{**}\equiv\omega_c(a/R_c)\\[0.2cm]
-\ T^{-1}\ B^{5/2}, & \qquad \qquad  T\gg \omega_c(a/R_c),
\end{array}
\right.
\end{equation}
The sign of the peak oscillates just like in the diffusive regime.

\textit{Ballistic regime}, $\omega_c/\Delta \ll R_c/a \ll N\Delta/\omega_c$.
This is the regime we have studied in the main part of the paper.
For the reader's convenience, we repeat the results here. The peak is governed
by the ${\cal O}(\Delta/\omega_c)$-contribution, its sign oscillates
and is positive for matching densities,
\begin{equation}
\label{peak-mod-ball-scaling}
\rho_{xx}^D \propto \left\{
\begin{array}{ll}
 T^2\ B\ \ln(B_*/B), & \qquad \qquad T\ll \Delta, \\[0.2cm]
 T^{-3}\ B^{7/2}\ \ln(B_*/B), & \qquad \qquad \Delta \ll T \ll T_*,\\[0.2cm]
 -\ T^{-1}\ B^{5/2}, & \qquad \qquad  T\gg T_*.
\end{array}
\right.
\end{equation}

\textit{Ultra-ballistic regime}, $N\Delta/\omega_c\ll R_c/a$.
The drag for all temperatures is determined by the
conventional ${\cal O}(q/k_F)$-contribution and is always positive,
\begin{equation}
\label{peak-ultra-ball-scaling}
\rho_{xx}^D \propto \left\{
\begin{array}{ll}
T^2\ B^2, & \qquad \qquad T\ll \Delta, \\[0.2cm]
T^{-1}\ B^{7/2}, & \qquad \qquad  T\gg \Delta,
\end{array}
\right.
\end{equation}

At high temperature, $T\gg \omega_c,$ the drag is governed by the conventional
contribution (and is therefore positive) in all the regimes.
It is linear in $T$ in the diffusive regime ($\rho_{xx}^D\propto T B^{-1/2}$).
In all the ballistic regimes the drag resistivity scales as
$\rho_{xx}^D \propto T^2 B^{1/2}$ for $\omega_c \ll T\ll v_F/a$
and $ \rho_{xx}^D \propto T B^{1/2}$ for $ T\gg v_F/a.$ (As mentioned in the end
of Sec.~\ref{highTdrag}, we do not consider
the magnetoplasmon contribution~\cite{Khaetskii}
here.)

\section{Summary}
\label{summary}

In this paper, we have developed a systematic diagrammatic theory of
the Coulomb drag in moderately strong magnetic fields, when the Landau
bands are already separated but the Landau level index is still
large. Using the self-consistent Born approximation, we performed a
thorough analysis of all relevant contributions and, on
this basis, analyzed the temperature dependence of the drag resistivity.
Depending on the relation between the cyclotron radius $R_c$
and the interlayer distance $a$ we distinguish several regimes.
We concentrated on the experimentally most relevant ballistic
regime. In this case the theoretical analysis requires  special care,
in view of a cancellation between leading-order contributions to the
triangle vertex ${\bf \Gamma}$. We also briefly considered the
evolution of the drag resistivity in the whole range of $R_c/a$, from
the diffusive to the ultra-ballistic regime.
\begin{figure}
\centering
\epsfxsize8cm\epsfbox{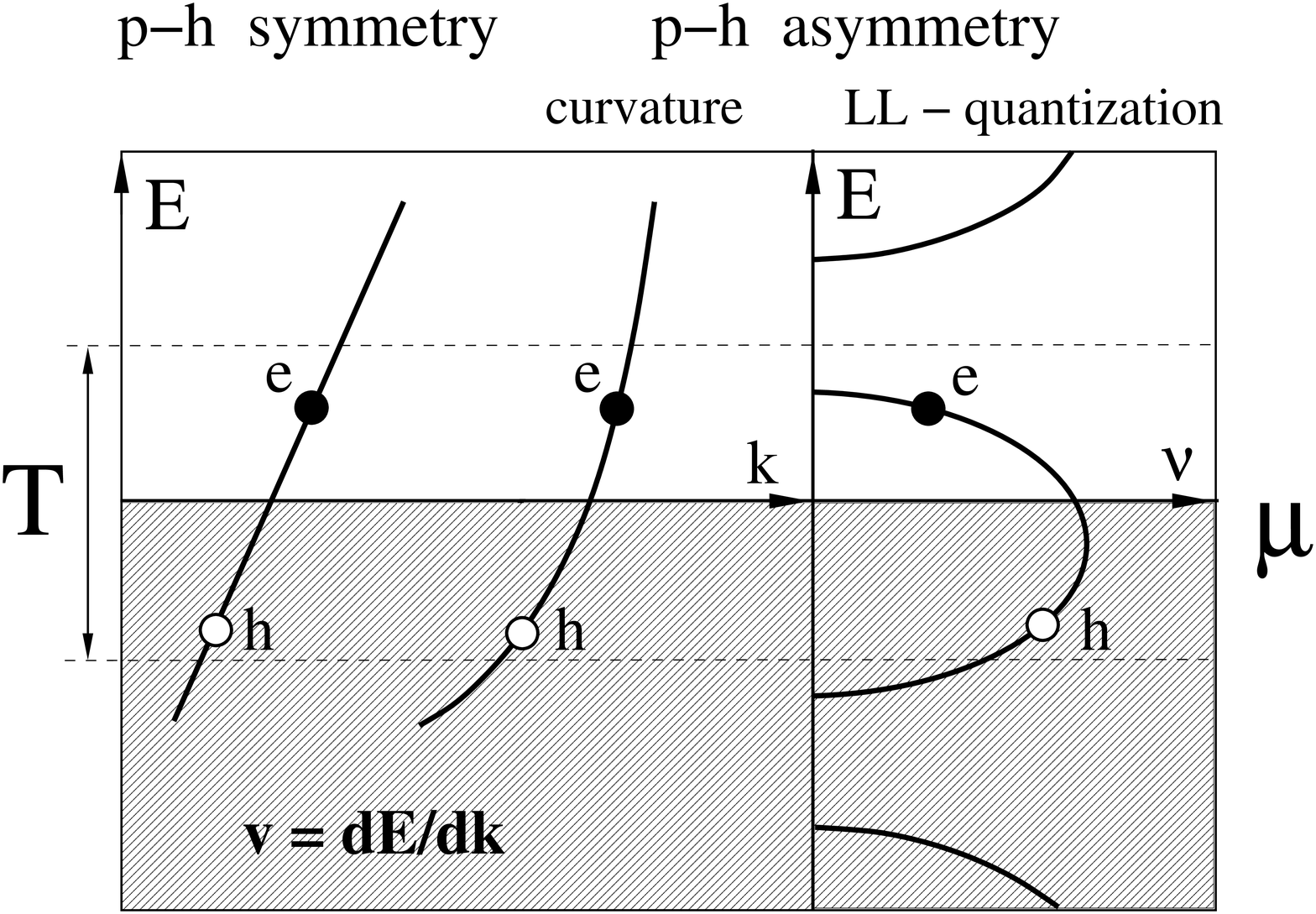}\hskip1cm
\epsfxsize7.8cm\epsfbox{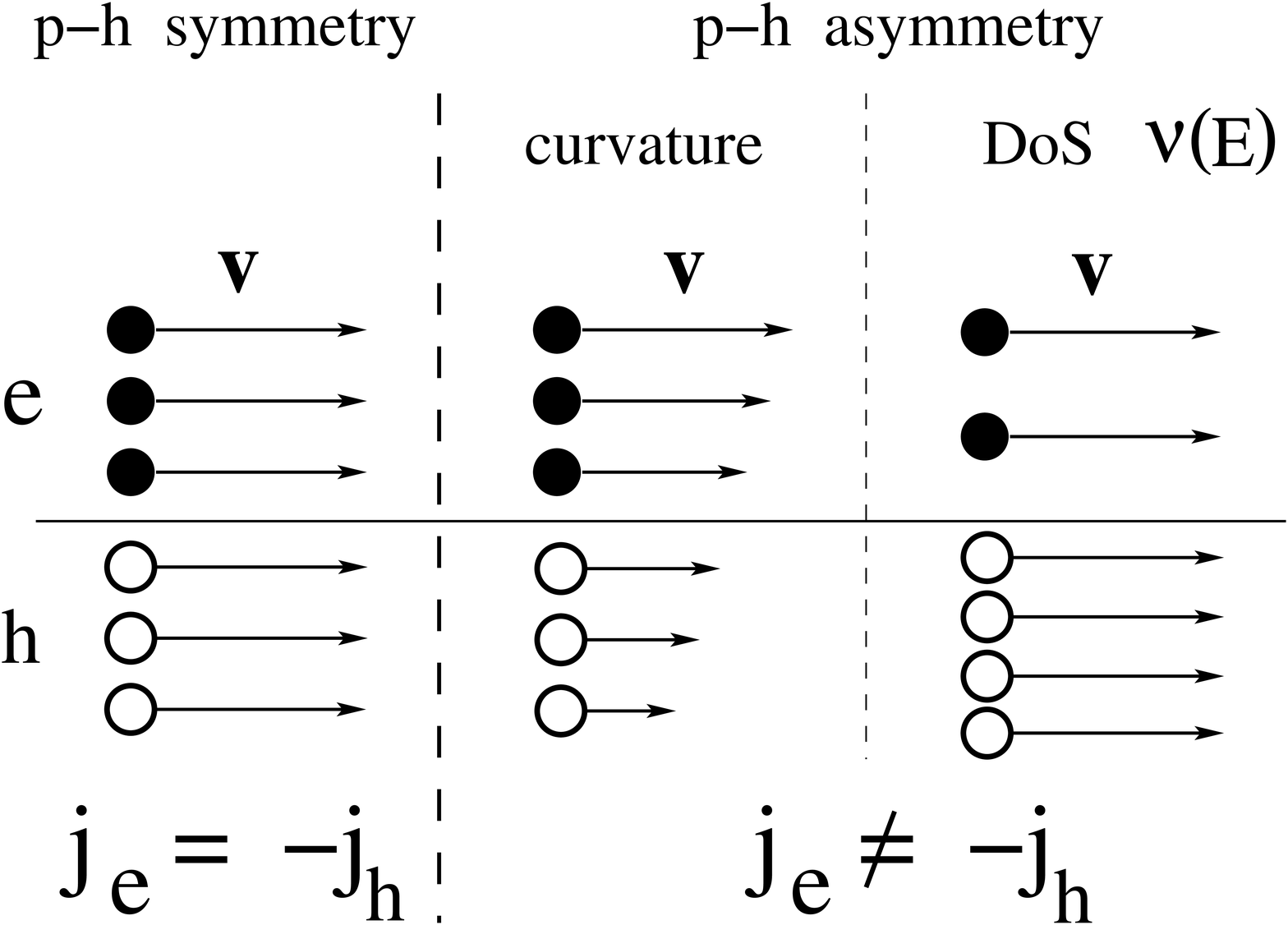}\vskip .3truecm
\caption{Schematic illustration of different sources of particle-hole asymmetry:
curvature of zero-$B$ spectrum $E(k)$ vs LL-quantization of the density of states (DoS)
$\nu(E)$.
In the particle-hole (p-h) symmetric case, the electronic and hole contributions
to the current induced in the passive layer ($j_e$ and
$j_h$, respectively) compensate each other.
When the p-h asymmetry is generated by a finite curvature, the velocities
of electrons and holes (shown by arrows in the right panel) are different,
which destroys the compensation. This is the ``conventional'' mechanism of the drag.
When the DoS depends on energy (in the present case because of the LL-quantization), an
``anomalous'' drag arises due to the difference in numbers of occupied electronic and hole states.
}
\label{fig-ph}
\end{figure}

We have shown that Coulomb drag in strong magnetic fields is an interplay
of two contributions, as illustrated in Fig.~\ref{fig-ph}.
At high temperatures, the leading contribution
is due to breaking of particle-hole symmetry by the curvature
of the zero-$B$ electron spectrum. This ``normal'' contribution to the
drag is always positive and increases in a broad temperature range as
$T^2$. At low temperatures, we find that a second,  ``anomalous'',
contribution dominates, which arises from the breaking of
particle-hole symmetry by the energy dependence of the density of
states related to Landau quantization. This contribution is sharply
peaked at a temperature $T\sim\Delta$ (where $\Delta$ is the Landau
level width) and has an oscillatory sign depending on the density
mismatch between the two layers. In particular, we find that in the
ballistic regime the sign is positive for equal densities, in contrast
to the negative sign in the diffusive regime found in Ref.\ \onlinecite{Oppen}.

Our results for the temperature dependence and sign of the drag
resistivity $\rho_{xx}^D(T)$ in the ballistic regime
are illustrated in Fig.~\ref{fig6}. These results are in good
agreement with recent experimental findings \cite{Muraki}, and thus
explain the remarkable features of Coulomb drag in high Landau
levels observed experimentally.

Finally, we discuss some prospects for future research.
First, our
theory can be generalized to phonon drag, which is expected to
dominate over Coulomb drag at larger separations between the
layers. Second, it will be interesting to consider the magnetic field
and temperature dependence of the drag around filling factor
$\nu=1/2$, where transport is due to composite fermions moving in
a reduced magnetic field.\cite{foot-CF}
Third, one can study the effects of quantum
localization, as well as criticality in the center of the Landau
band,\cite{Shimshoni}
which should become important in lower Landau levels or for very low
temperatures. Finally, it should be possible to reproduce our
results within the framework of a quantum kinetic equation
[cf. Ref.~\onlinecite{Vavilov}].
This would also allow one to generalize the theory
of magnetodrag to non-equilibrium setups
(strong bias, microwave, etc.), as well as to other observables (e.g., the
thermopower) related to particle-hole asymmetry.

\acknowledgments

We are grateful to K.~von~Klitzing, J.G.S.~Lok, and K.~Muraki
for informing us on experimental results prior to publication
and for interesting discussions.
We further acknowledge valuable discussions with I.L.~Aleiner,
J.~Dietel,
A.V.~Khaetskii, and A.~Stern.
FvO thanks the Weizmann Institute for hospitality and support
through the Einstein Center and LSF
while part of this work was performed.
Financial support by the DFG-Schwerpunktprogramm
``Quanten-Hall-Systeme" (IVG, ADM, and FvO),
by SFB 290 and the ``Junge Akademie" (FvO)
is gratefully acknowledged.

\appendix

\section{Analytical continuation}
\label{app:analytical}

\begin{figure}
\centering
\epsfxsize10cm\epsfbox{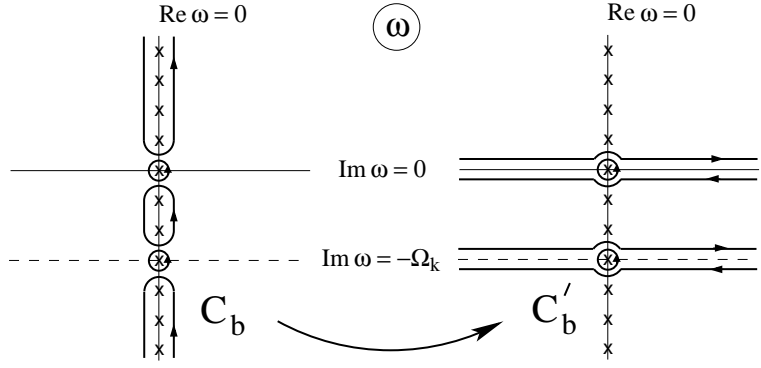} \vskip .3truecm
\caption{Contours for the $\omega$-integration.
 }
\label{fig8}
\end{figure}

In this Appendix we perform the analytical continuation of the Matsubara
expressions for the drag conductivity and the triangle vertex $\bf\Gamma$.
To calculate the Matsubara sum over $\omega_n=2\pi n T$ in
Eq.~(\ref{drag-conductivity-matsubara}), the standard contour
integration in the complex $\omega$ plane is done,
\begin{equation}
T\sum_{\omega_n}f(i\omega_n)={1\over 4\pi i} \int_{C_b} d\omega\ f(\omega)
\ {\rm coth}{\omega\over 2 T}.
\label{analyt-boson-main}
\end{equation}
The integrand has branch cuts at ${\rm Im}\ \omega=0$ and
${\rm Im}\ \omega=-\Omega_k,$
where $\Omega_k$ represents the external frequency. The integration contour $C_b$ thus
contains three parts, see Fig.~\ref{fig8} Deforming the contour as shown in Fig.~\ref{fig8},
we get four
terms corresponding to four
lines (above and below of both the branch cuts)
forming the new contour,
\begin{eqnarray}
\label{drag-matsubara-four}
\sigma_{ij}^{D}(i\Omega_k)&=&
-{e^2\over 8 \Omega_k S}\sum_{\bf q}
\int_{-\infty}^{\infty} d\omega\   {\rm coth}{\omega\over 2 T}
\nonumber \\
&\times& \left[
\Gamma_i^{(1)}({\bf q},\omega+i\Omega_k,\omega+i0)
\Gamma_j^{(2)}({\bf q},\omega+i0,\omega+i\Omega_k)
U({\bf q},\omega+i\Omega_k)U({\bf q},\omega+i0)
\right. \nonumber \\
&-&
\Gamma_i^{(1)}({\bf q},\omega+i\Omega_k,\omega-i0)
\Gamma_j^{(2)}({\bf q},\omega-i0,\omega+i\Omega_k)
U({\bf q},\omega+i\Omega_k)U({\bf q},\omega-i0)
\nonumber \\
&+&
\Gamma_i^{(1)}({\bf q},\omega+i0,\omega-i\Omega_k)
\Gamma_j^{(2)}({\bf q},\omega-i\Omega_k,\omega+i0)
U({\bf q},\omega+i0)U({\bf q},\omega-i\Omega_k)
\nonumber \\
&-&\left.
\Gamma_i^{(1)}({\bf q},\omega-i0,\omega-i\Omega_k)
\Gamma_j^{(2)}({\bf q},\omega-i\Omega_k,\omega-i0)
U({\bf q},\omega-i0)U({\bf q},\omega-i\Omega_k)
\right].
\end{eqnarray}
In the third and fourth terms
we have used ${\rm coth}(z+i\Omega_k/2T)={\rm coth} z$.
The contributions of points $\omega=0$ and $\omega=-i\Omega_k$
cancel the integral over the small circles around these points, so that the
integrals above should be understood in the principal value sense.

We now perform the analytical continuation $i\Omega_k\to\Omega+i0$ and
take the limit $\Omega\to 0$.
As shown in Ref.~\onlinecite{Kamenev}, the first and the last terms coming from
outer sides of branch cuts
vanish in the limit $\Omega\to 0$.
This yields
\begin{eqnarray}
\label{drag-continued}
\sigma_{ij}^{D}&=&-{e^2\over 8\pi S}\sum_{{\bf q}}
\int_{-\infty}^\infty d\omega\  {\rm coth}{\omega\over 2 T}\
\frac{\partial}{\partial\omega}\nonumber \\
&\times&
\left[
\Gamma_i^{(1)}({\bf q},\omega+i0,\omega-i0)\Gamma_j^{(2)}({\bf q},\omega-i0,\omega+i0)
U({\bf q},\omega+i0)U({\bf q},\omega-i0)
 \right].
\end{eqnarray}
Using
\begin{equation}
\frac{\partial}{\partial\omega}{\rm coth}{\omega\over 2 T}
=-{1\over 2T {\rm sinh}^2(\omega/2T)},
\label{coth-sinh}
\end{equation}
we arrive at Eq.~(5).

\begin{figure}
\centering
\epsfxsize10cm\epsfbox{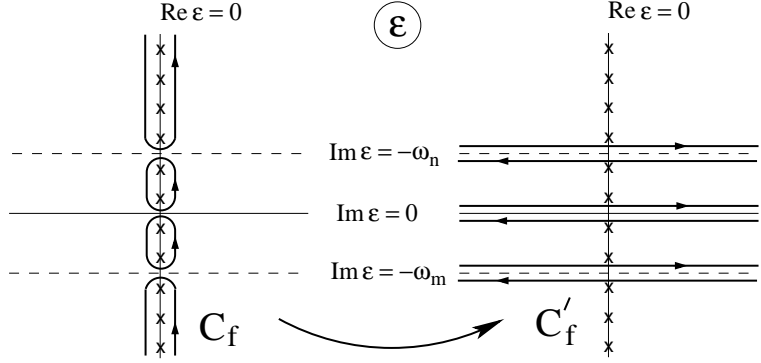} \vskip .3truecm
\caption{Contours for the $\epsilon$-integration.
 }
\label{fig9}
\end{figure}

The next step is the analytical continuation of the triangle vertex.
The summation over the fermionic Matsubara energies
$\epsilon_k=2\pi(k+1/2 T$ in Eq.~(\ref{triangle-matsubara})
is performed using the integral
\begin{equation}
T\sum_{\epsilon_k}f(i\epsilon_k)={1\over 4\pi i} \int_{C_f} d\epsilon f(\epsilon)
\ {\rm tanh}{\epsilon\over 2 T},
\label{analyt-fermi-main}
\end{equation}
along the contour $C_f$ shown in Fig.~\ref{fig9}.
Since the triangle vertex depends on
two frequencies $i\omega_m$ and $i\omega_n$, the integrand now has three
branch cuts in the complex plane of $\epsilon$, namely at ${\rm Im}\ \epsilon=0,$
${\rm Im}\ \epsilon=-\omega_m,$ and ${\rm Im}\ \epsilon=-\omega_n.$
Similarly to $C_b$, the contour $C_f$ can be deformed into a set
of six lines going on both sides of each of the branch cuts (see Fig.~\ref{fig9}),
yielding
\begin{eqnarray}
{\bf\Gamma}({\bf q},i\omega_m,i\omega_n)&=& \int_{-\infty}^\infty
{d\epsilon \over 4\pi i} \tanh{\epsilon \over 2T}
\nonumber \\
    &\times&{\rm tr}
\left\{
{\bf v}\
[{\cal G}^+(\epsilon)-{\cal G}^-(\epsilon)]
e^{i{\bf qr}}
{\cal G}(\epsilon-i\omega_n)
e^{-i{\bf qr}}
{\cal G}(\epsilon+i\omega_m-i\omega_n)\right. \nonumber \\
     &-&
{\bf v}\ {\cal G}(\epsilon+i\omega_n)
e^{i{\bf qr}}
[{\cal G}^+(\epsilon)-{\cal G}^-(\epsilon)]
e^{-i{\bf qr}}
{\cal G}(\epsilon+i\omega_m) \nonumber   \\
     &+&
\left.
{\bf v}\ {\cal G}(\epsilon-i\omega_m+i\omega_n)
e^{i{\bf qr}}
{\cal G}(\epsilon-i\omega_m)
e^{-i{\bf qr}}
[{\cal G}^+(\epsilon)-{\cal G}^-(\epsilon)]
\right\}
\nonumber \\
     &+&
(\omega_n \to -\omega_m,{\bf q}\to-{\bf q}).
\label{Gamma-summed}
\end{eqnarray}
In this formula ${\cal G}^\pm(\epsilon)={\cal G}(\epsilon\pm i0)$ and we have used
${\rm tanh}(z-i\omega_m/2T)={\rm tanh}(z-i\omega_n/2T)={\rm tanh}z$.
The equation (\ref{Gamma-summed}) is valid irrespective of the relation between
$\omega_m, \ \omega_n,$ and $0$. Performing the analytical continuation
to real frequencies $i\omega_m\to \omega_1+i0$ and  $i\omega_n\to \omega_2-i0$
(and shifting the integration variables
$\epsilon\to \epsilon+\omega_2$ and $\epsilon \to \epsilon+\omega_1$
in the first and third terms, respectively)
we obtain
\begin{eqnarray}
{\bf\Gamma}({\bf q},\omega_1+i0,\omega_2-i0)
&=& \int_{-\infty}^\infty
{d\epsilon \over 4\pi i}
\nonumber \\
    &\times&{\rm tr}
\left\{
\tanh{\epsilon+\omega_2 \over 2T}\
{\bf v}\
[{\cal G}^+(\epsilon+\omega_2)-{\cal G}^-(\epsilon+\omega_2)]
e^{i{\bf qr}}
{\cal G}^+(\epsilon)
e^{-i{\bf qr}}
{\cal G}^+(\epsilon+\omega_1)\right. \nonumber \\
     &-&
\tanh{\epsilon \over 2T}\
{\bf v}\ {\cal G}^-(\epsilon+\omega_2)
e^{i{\bf qr}}
[{\cal G}^+(\epsilon)-{\cal G}^-(\epsilon)]
e^{-i{\bf qr}}
{\cal G}^+(\epsilon+\omega_1) \nonumber   \\
     &+&
\tanh{\epsilon+\omega_1 \over 2T}
\left.
{\bf v}\ {\cal G}^-(\epsilon+\omega_2)
e^{i{\bf qr}}
{\cal G}^-(\epsilon)
e^{-i{\bf qr}}
[{\cal G}^+(\epsilon+\omega_1)-{\cal G}^-(\epsilon+\omega_1)]
\right\}
\nonumber \\
     &+&
(\omega,{\bf q}\to-\omega,-{\bf q}).
\label{Gamma-real-freq}
\end{eqnarray}
Setting $\omega_1=\omega_2$ and collecting the contributions containing
only retarded (from the first term) and only advanced (from the third term)
Green functions, we arrive [up to a redefinition of zero
of fermionic energies, which are counted from the chemical potential
in Eq.~(\ref{Gamma-real-freq})] at Eq.~(\ref{triangle(a)}) for ${\bf\Gamma}^{(a)}$.
The remaining terms constitute the expression
(\ref{triangle(b)}) for ${\bf\Gamma}^{(b)}$.

\section{Vertex corrections in SCBA}
\label{app:vertex}

\begin{figure}
\centering
\vskip1cm
\epsfxsize10cm\epsfbox{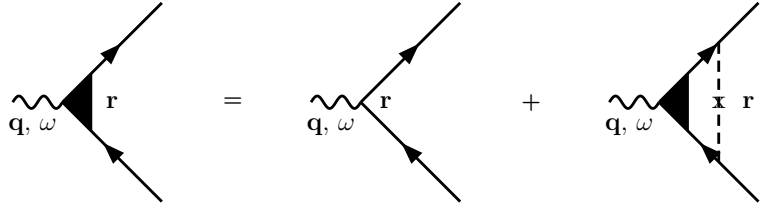} \vskip1cm
\caption{Diagrams for the (scalar) vertex corrections in real space.
 }
\label{fig10}
\end{figure}

In this appendix, we review vertex corrections in SCBA. We start by noting that in real space,
the impurity-averaged electron Green function in SCBA can be written as 
\begin{equation}
   G( {\bf r},{\bf r}';E) = e^{i\varphi({\bf r},{\bf r}')} \sum_n C_n( {\bf r}-{\bf r}') 
           G_n(E)
\end{equation}
with 
\begin{equation}
  C_n({\bf r},{\bf r}') = {1\over 2\pi\ell^2} e^{-({\bf r}-{\bf r}')^2/2\ell^2} L_n\left( ({\bf r}-{\bf r}')^2
     \over 2\ell^2\right).
\end{equation}
The gauge-dependent phase $\varphi({\bf r},{\bf r}')$
satisfies $\varphi({\bf r},{\bf r}') =-\varphi({\bf r}',{\bf r})$.
This can be used to express the vertex correction in real space
as (cf.\ Fig.\ \ref{fig10}) 
\begin{equation}
  \gamma^{\mu\nu}({\bf q},\omega;{\bf r}) = e^{i{\bf qr}} + {1\over 2\pi \nu_0 \tau_0}
    \int d{\bf r}' \gamma^{\mu\nu}({\bf q},\omega; {\bf r}')
    G^\mu({\bf r},{\bf r}';E+\omega) G^\nu({\bf r}',{\bf r};E).
\end{equation}
For well-separated Landau levels, the valence LL with LL index N gives the dominant contribution so that
\begin{equation}
  \gamma^{\mu\nu}({\bf q},\omega;{\bf r}) = e^{i{\bf qr}} + {1\over 2\pi \nu_0 \tau_0}
  G_N^\mu(E+\omega) G_N^\nu(E)
    \int d{\bf r}' C_N({\bf r}-{\bf r}')C_N({\bf r}'-{\bf r})\gamma^{\mu\nu}({\bf q},\omega; {\bf r}').
\end{equation}
Thus, we find that
\begin{equation}
   \gamma^{\mu\nu}( {\bf q} , \omega ; {\bf r}) = \gamma^{\mu\nu}( {\bf q} , \omega) e^{i{\bf qr}}
\end{equation}
with
\begin{eqnarray}
     \gamma^{\mu\nu}( {\bf q} , \omega) = 1  +  {(2\pi\ell^2)\Delta^2\over 4}
      \gamma^{\mu\nu}({\bf q},\omega)
      G_N^\mu(E+\omega) G_N^\nu(E)
      \int d{\bf r}' C_N({\bf r}-{\bf r}')C_N({\bf r}'-{\bf r}) e^{-i{\bf q}({\bf r}-{\bf r}')}.
\end{eqnarray}
Here we used the identity $1/2\pi\nu_0\tau_0=(2\pi\ell^2)\Delta^2/4$.
The integral is equal to
\begin{eqnarray}
   \int d{\bf r}' C_N({\bf r}-{\bf r}')C_N({\bf r}'-{\bf r}) e^{-i{\bf q}({\bf r}-{\bf r}')}
       &=& {1\over2\pi\ell^2}e^{ q^2\ell^2}[L_N\left( {q^2\ell^2/ 2}  \right)]^2
        \nonumber\\
       &\simeq& {1\over2\pi\ell^2}J_0^2(qR_c),
\end{eqnarray}
where the second equality holds in the limit of high Landau levels. Neglecting the
frequency-dependence and using the identities
\begin{eqnarray}
   G_N^+G_N^- &=& {4\over \Delta^2} \\
   G_N^+G_N^+ &=& {1\over (\Sigma_N^-)^2},
\end{eqnarray}
we can solve for $\gamma^{\mu\nu}$, and obtain Eqs.\ (\ref{scba-vertex++}) and (\ref{scba-vertex+-})
for the vertex corrections.
Finally, for finite $\omega$ we get
\begin{equation}
\gamma^{\mu\nu}({\bf q} , \omega)=\frac{1}{1-(\Delta^2/4)J_0^2(qR_c)G_N^\mu(E+\omega) G_N^\nu(E)}
,
\end{equation}
which is used in Eq.~(\ref{leading-gamma-finiteTw}).

\section{Corrections of order $\Delta/\omega_c$}
\label{app:gamma}

In this appendix, we consider the contributions to the triangle vertex to order $\Delta/\omega_c$
in more detail. To this order, vertex corrections of the scalar vertices can be neglected.

We first consider the case (i) in which both Green functions adjacent to the current vertex are evaluated
in Landau levels other than $N$. As mentioned in Sec.\ \ref{llmixing}, the Green function connecting the scalar
vertices should be evaluated in the $N$th LL up to corrections of order $(\Delta/\omega_c)^2$.
Using the semiclassical expression (\ref{matr-el-quasicl})
for the matrix elements, we then find for the corresponding
correction to ${\bf \Gamma}^{(b)}$ the expression
\begin{eqnarray}
   \delta{\bf \Gamma}^{(b)}({\bf q},\omega)&=& -{\omega\over i\pi} {1\over 2\pi\ell^2}
     {\sqrt{N}\over \ell m \sqrt{2}}
     (G_N^+-G_N^-)\sum_{n\neq N,N+1}\left( 2i\atop 0\right) {1\over (N-n) \omega_c}{1\over (N-n+1)\omega_c}
       iJ_{N-n+1}(qR_c)J_{N-n}(qR_c)
    \nonumber\\
    &=& -{\omega\over  2\pi^2\ell^2} {\sqrt{N}\over \ell m \sqrt{2}}
     {(G_N^+-G_N^-) \over \omega_c^2}  \sum_{k=1}^\infty{1\over k(k+1)}\left( 2i\atop 0\right)
       \left[J_{k}(qR_c)J_{k+1}(qR_c)
      +J_{-k}(qR_c)J_{-(k+1)}(qR_c)\right].
\end{eqnarray}
Using that $J_{-k}(z)=(-1)^k J_k(z)$, we find that the expression in square brackets vanishes
so that $\delta{\bf \Gamma}^{(b)}({\bf q},\omega)=0$.

The corresponding contribution to ${\bf \Gamma}^{(a)}({\bf q},\omega)$ takes the form
\begin{eqnarray}
   \delta{\bf \Gamma}^{(a)}({\bf q},\omega) &=& {\omega\over \pi} {1\over 2\pi\ell^2}
      {\rm Im} {\partial\over\partial {\bf q}} {\sum_{n,m}}^\prime J_{n-m}^2(qR_c) G_m^+G_n^+
       \nonumber\\
       &=& {2\omega\over \pi} {1\over 2\pi\ell^2}
      {\rm Im} {\partial\over\partial {\bf q}} {\sum_{k=1}^\infty} J_{k}^2(qR_c)
      (G_N^+G_{N-k}^+ + G_N^+G_{N+k}^+).
      \label{C2}
\end{eqnarray}
The prime on the sum indicates that only those terms should be kept, in which one of the two
Green functions is evaluated in a LL different from $N$. In leading order,
$$ G_{N-k}^+=-G_{N+k}^+={1\over k\omega_c},$$
and hence also the contribution (\ref{C2}) to the triangle vertex vanishes.

Next, we turn to the contribution (ii) in which the diagrams in Fig.\ \ref{fig4} are evaluated to
next-to-leading order in $\Delta/\omega_c$
while neglecting the vertex corrections on the scalar vertices. Such contributions can arise in
particular from the self energy entering
$G_{N\pm 1}$. We first consider the corresponding contribution to ${\Gamma}^{(b)}_x$.
(For the purpose of this appendix, we choose ${\bf q}\parallel {\bf\hat x}$.)
According to the diagrams in Fig.\ \ref{fig4}, we have for the contribution (ii)
\begin{equation}
    \Gamma^{(b)}_x =
    {\omega\over \pi} {1\over 2\pi\ell^2} {\sqrt{N} \over \ell m \sqrt{2}} J_0(qR_c)J_1(qR_c)
       2{\rm Im} (G_{N-1}^- - G_{N+1}^-)[G_N^+]^2.
\end{equation}
Here, we have already used that to the order under consideration,
\begin{equation}
  -(G^-_{N-1}-G^-_{N+1})G_N^+G_N^- + (G_{N-1}^+ - G_{N+1}^+) G_N^+G_N^- \simeq 0.
\end{equation}
To our order, we then find
\begin{equation}
    \Gamma^{(b)}_x =
    {\omega\sqrt{2N}\over \pi^2\ell}  J_0(qR_c)J_1(qR_c)
       {\rm Im}[G_N^+]^2.
\end{equation}
Comparing with Eq.\ (\ref{leading-gammaa}), we find
even to order $\Delta/\omega_c$ that this contribution
is cancelled exactly by $\Gamma^{(a)}_x$. Thus,
there is also no contribution of type (ii) to
$\Gamma_x$ and $\Gamma_x$ vanishes to order $\Delta/\omega_c$.

Finally, we consider the contribution of type (ii) to $\Gamma_y$. Since this is a transverse
contribution, we need to consider only ${\bf \Gamma}^{(b)}$. In this case, the diagrams
in Fig.\ \ref{fig4} translate into the expression
\begin{equation}
\label{appgammay}
  \Gamma_y = {i\omega\over \pi} {1\over 2\pi\ell^2} {\sqrt{N} \over \ell m\sqrt{2}}
    iJ_0(qR_c)J_1(qR_c) (G_N^+-G_N^-)2i{\rm Im} (G_{N-1}^-+G_{N+1}^-)G_N^+
\end{equation}
Noting that the leading order cancels from the combination, $G_{N-1}^-+G_{N+1}^-$, we can
simply evaluate Eq.\ (\ref{appgammay}) for $\Gamma_y$ to leading nonvanishing order. This yields
Eq.\ (\ref{gamma-llmixing}) for ${\bf \Gamma}^{(\Delta/\omega_c)}$ in the main text.

\section{
Contributions to drag from different momentum regions}
\label{app:drag-split}

We write down explicitly the momentum integrals determining the function $I(\omega)$
in Eq.~(\ref{rho-w-integral}).
The first integral, corresponding to the diffusive range of momenta $qR_c\ll 1,$
\begin{eqnarray}
I_{\rm I}(\omega)&=&-\int_0^{1/R_c}{dq\ q\over 2\pi}
\left(\frac{q}{\sinh q a}\right)^2
\left\{4qR_c \frac{(\mu-E_N)D(\mu) q^2}{[D(\mu) q^2]^2+\omega^2}\right\}_1
\left\{4qR_c \frac{(\mu-E_N)D(\mu) q^2}{[D(\mu) q^2]^2+\omega^2}\right\}_2
\nonumber \\
&\times& \left\{\frac{\Delta^2}{\Delta^2-(\mu-E_N)^2}
\left(\pi \Delta\over 2\omega_c\right)^2 \frac{[D(\mu) q^2]^2+\omega^2}{[D(\mu) q^2]^2} \right\}_1
\left\{\frac{\Delta^2}{\Delta^2-(\mu-E_N)^2}
\left(\pi \Delta\over 2\omega_c\right)^2 \frac{[D(\mu) q^2]^2+\omega^2}{[D(\mu) q^2]^2} \right\}_2\ ,
\end{eqnarray}
is dominated by the contribution of the ``diffusive rectification'', Eq.~(\ref{diffusive-omega}),
while the screening is determined by Eq.~(\ref{diffusive-polarization}).
The second integral
\begin{equation}
I_{\rm II}(\omega)=I_{\rm II-1}(\omega)+I_{\rm II-2}(\omega),
\label{I-II}
\end{equation}
includes the contribution of $\Gamma^{(1/qR_c)}$
[denoted by $I_{\rm II-1}(\omega)$] and $\Gamma^{(\Delta/\omega_c)}$
[denoted by $I_{\rm II-2}(\omega)$], Eqs.~(\ref{gamma-1/qrc}) and (\ref{gamma-llmixing}), respectively,
while the screening in $I_{\rm II}$ is determined by $N$th LL, Eq.~(\ref{ballistic-realpolarization}).
\begin{eqnarray}
I_{\rm II-1}(\omega)&=&
-\int_{1/R_c}^{\omega_c/\Delta R_c}{dq\ q\over 2\pi}
\left(\frac{q}{\sinh q a}\right)^2\
\left\{J_1(qR_c)J_0^3(qR_c)\right\}_1 \
\left\{J_1(qR_c)J_0^3(qR_c)\right\}_2
\nonumber \\
&\times&
\left\{\frac{64}{\Delta^4} (\mu-E_N)[\Delta^2-(\mu-E_N)^2]^{3/2}\right\}_1
\left\{\frac{64}{\Delta^4} (\mu-E_N)[\Delta^2-(\mu-E_N)^2]^{3/2}\right\}_2
\nonumber \\
&\times&
\left\{1+{8\omega_c\over 3\pi\Delta}J_0^2(qR_c)
\left[1-{(\mu-E_N)^2\over \Delta^2}\right]^{3/2}\right\}^{-2}_1
\left\{1+{8\omega_c\over 3\pi\Delta}J_0^2(qR_c)
\left[1-{(\mu-E_N)^2\over \Delta^2}\right]^{3/2}\right\}^{-2}_1\ ,
 \label{I-II-1} \\
I_{\rm II-2}(\omega)&=&
\int_{1/R_c}^{\omega_c/\Delta R_c}{dq\ q\over 2\pi}
\left(\frac{q}{\sinh q a}\right)^2 \
\left\{J_1(qR_c)J_0(qR_c)\right\}_1 \
\left\{J_1(qR_c)J_0(qR_c)\right\}_2
\nonumber \\
&\times&
\left\{\frac{16}{\omega_c\Delta^2} (\mu-E_N)[\Delta^2-(\mu-E_N)^2]\right\}_1
\left\{\frac{16}{\omega_c\Delta^2} (\mu-E_N)[\Delta^2-(\mu-E_N)^2]\right\}_2
\nonumber \\
&\times&
\left\{1+{8\omega_c\over 3\pi\Delta}J_0^2(qR_c)
\left[1-{(\mu-E_N)^2\over \Delta^2}\right]^{3/2}\right\}^{-2}_1
\left\{1+{8\omega_c\over 3\pi\Delta}J_0^2(qR_c)
\left[1-{(\mu-E_N)^2\over \Delta^2}\right]^{3/2}\right\}^{-2}_1.
\label{I-II-2}
\end{eqnarray}
The integration domain in the third integral, $I_{\rm III}(\omega),$
corresponds to the range where the screening acquires its static zero-$B$ form, Eq.~(\ref{nLL-Pi}),
while the triangle vertex is dominated by $\Gamma^{(\Delta/\omega_c)}$, Eq.~(\ref{gamma-llmixing}),
\begin{eqnarray}
I_{\rm III}(\omega)&=&\int_{\omega_c/\Delta R_c}{dq\ q\over 2\pi}
\left(\frac{q}{\sinh q a}\right)^2\{J_1(qR_c)J_0(qR_c)\}_1 \{J_1(qR_c)J_0(qR_c)\}_2
\nonumber \\
&\times&
\left\{\frac{16}{\omega_c\Delta^2} (\mu-E_N)[\Delta^2-(\mu-E_N)^2]\right\}_1
\left\{\frac{16}{\omega_c\Delta^2} (\mu-E_N)[\Delta^2-(\mu-E_N)^2]\right\}_2
\label{I-III}
\end{eqnarray}

Let us analyze the first term in $I_{\rm II}$, Eq.~(\ref{I-II-1}).
Consider identical layers.
The screening is nontrivial and almost vanishes in the
vicinity of zeroes $Q_n$ of $J^2_0(qR_c)$.
The structure of the integral is
\begin{equation}
I_{\rm II-1}\propto \int_{1/R_c}^{\omega_c/\Delta R_c}
dq q \frac{J_1^2(qR_c)J_0^6(qR_c)}{[1+A J_0^2(qR_c) ]^4},
\label{I_II-1-simple}
\end{equation}
where $A\sim \omega_c/\Delta$.
We see that the integral is dominated by the momenta close to $Q_n$,
each peak contributing $\sim R_c^{-2} Q_n^{1/2} A^{-7/2}$,
so that the total result
\begin{equation}
I_{\rm II-1}\propto {1\over R_c^2}\sum_n  Q_n^{1/2}
\left({\omega_c\over \Delta}\right)^{-7/2}
\sim  {1\over R_c^2}\left({\Delta\over \omega_c}\right)^{7/2}
\int_1^{\omega_c/\Delta}dQ Q^{1/2}\sim
 {1\over R_c^2} \left({\Delta\over \omega_c}\right)^2,
\end{equation}
is determined by the upper limit where $A J^2_1(q R_c)\sim 1.$

Similarly, we estimate the second term in $I_{\rm II}$, Eq.~(\ref{I-II-2}),
\begin{equation}
\label{I_II-2-simple}
I_{\rm II-2}\propto  \left({\Delta\over \omega_c}\right)^2
\int_{1/R_c}^{\omega_c/\Delta R_c}dq q \frac{J_1^2(qR_c)J_0^2(qR_c)}{[1+A J_0^2(qR_c) ]^4}
\sim {1\over R_c^2} \left({\Delta\over \omega_c}\right)^{9/2}
\int_1^{\omega_c/\Delta}dQ Q^{3/2}\sim
{1\over R_c^2} \left({\Delta\over \omega_c}\right)^2,
\end{equation}
yielding the result of the same order as for Eq.~(\ref{I-II-1}), since both integrals
are dominated by the upper limit.
We note that for this reason the same estimate can be obtained by replacing
$J_0^2(qR_c),J_1^2(qR_c)$ by $(\pi q R_c)^{-1}.$
The two terms $I_{\rm II-1}$ and $I_{\rm II-2}$ give contributions
of the opposite signs to the drag resistivity,
since ${\cal O}(1/qR_c)\leftrightarrow \Gamma_{||}(B)=\Gamma_{||}(-B)$, while
 ${\cal O}(\Delta/\omega_c) \leftrightarrow \Gamma_{\perp}(B)=-\Gamma_{\perp}(-B)$.

Estimating other terms, we obtain
\begin{eqnarray}
I_{\rm I}&\sim& {1\over a^2 R_c^2}\left({\Delta\over \omega_c}\right)^4
\int_{Q_{\rm min}}^1{dQ\over Q}
= {1\over a^2 R_c^2}\left({\Delta\over \omega_c}\right)^4 \ln Q_{\rm min},
\label{I-I-result} \\
I_{\rm II}&\sim& {1\over a^2 R_c^2}
\left({\Delta\over \omega_c}\right)^{7/2}\int_1^{\omega_c/\Delta}dQ Q^{1/2}
\sim {1\over a^2R_c^2} \left({\Delta\over \omega_c}\right)^2,
\label{I-II-result} \\
I_{\rm III}&\sim& {1\over a^2 R_c^2}
\left({\Delta\over \omega_c}\right)^2
\int_{\omega_c/\Delta}^{R_c/a}{dQ\over Q}
= {1\over a^2 R_c^2} \left({\Delta\over \omega_c}\right)^2
\ln\left({R_c \Delta \over a \omega_c}\right),
\label{I-III-result}
\end{eqnarray}
where in the diffusive term $I_{\rm I}$ the momentum integration
is restricted from below by $Q_{\rm min} = R_c (\omega/\kappa_0 a D )^{1/2}
\sim R_c (T/\kappa_0 a D )^{1/2}.$ This infrared cut-off is necessary, since the
momentum integral diverges logarithmically at small $q$ in the diffusive regime, when
Eq.~(\ref{U12-simple}) and Eq.~(\ref{diffusive-polarization}) are used for the
interlayer interaction. The divergence
is naturally cured when the general formula (\ref{U12-general}) is employed
together with Eq.~(\ref{diffusive-polarization}).

Thus we conclude that at low temperatures $T\ll \Delta$ the total
integral is dominated by the
contribution of high momenta, $I_{\rm III},$
\begin{eqnarray}
I_{\rm I}+I_{\rm II}+I_{\rm III} &\simeq& I_{\rm III} =
{1\over 2 \pi^3 a^2 R_c^2} \ln\left({R_c \Delta \over a \omega_c}\right)
\nonumber \\
&\times&
\left\{\frac{16}{\omega_c\Delta^2} (\mu-E_N)[\Delta^2-(\mu-E_N)^2]\right\}_1
\left\{\frac{16}{\omega_c\Delta^2} (\mu-E_N)[\Delta^2-(\mu-E_N)^2]\right\}_2,
\label{sum3I}
\end{eqnarray}
resulting in Eq.~(\ref{rho-leading}).

\begin{figure}
\centering
\epsfxsize4.8cm\epsfbox{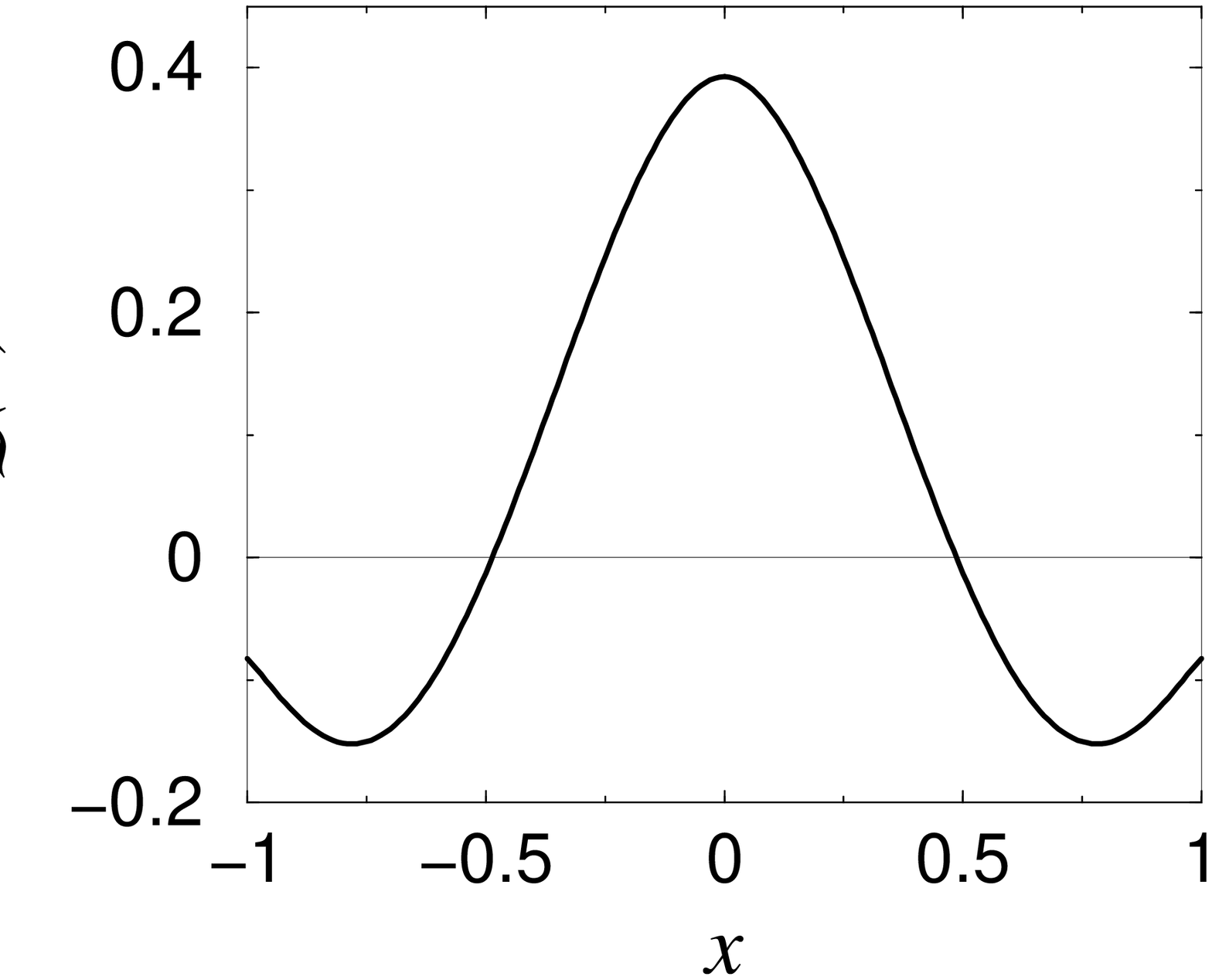}
\hskip0.3cm
\epsfxsize5cm\epsfbox{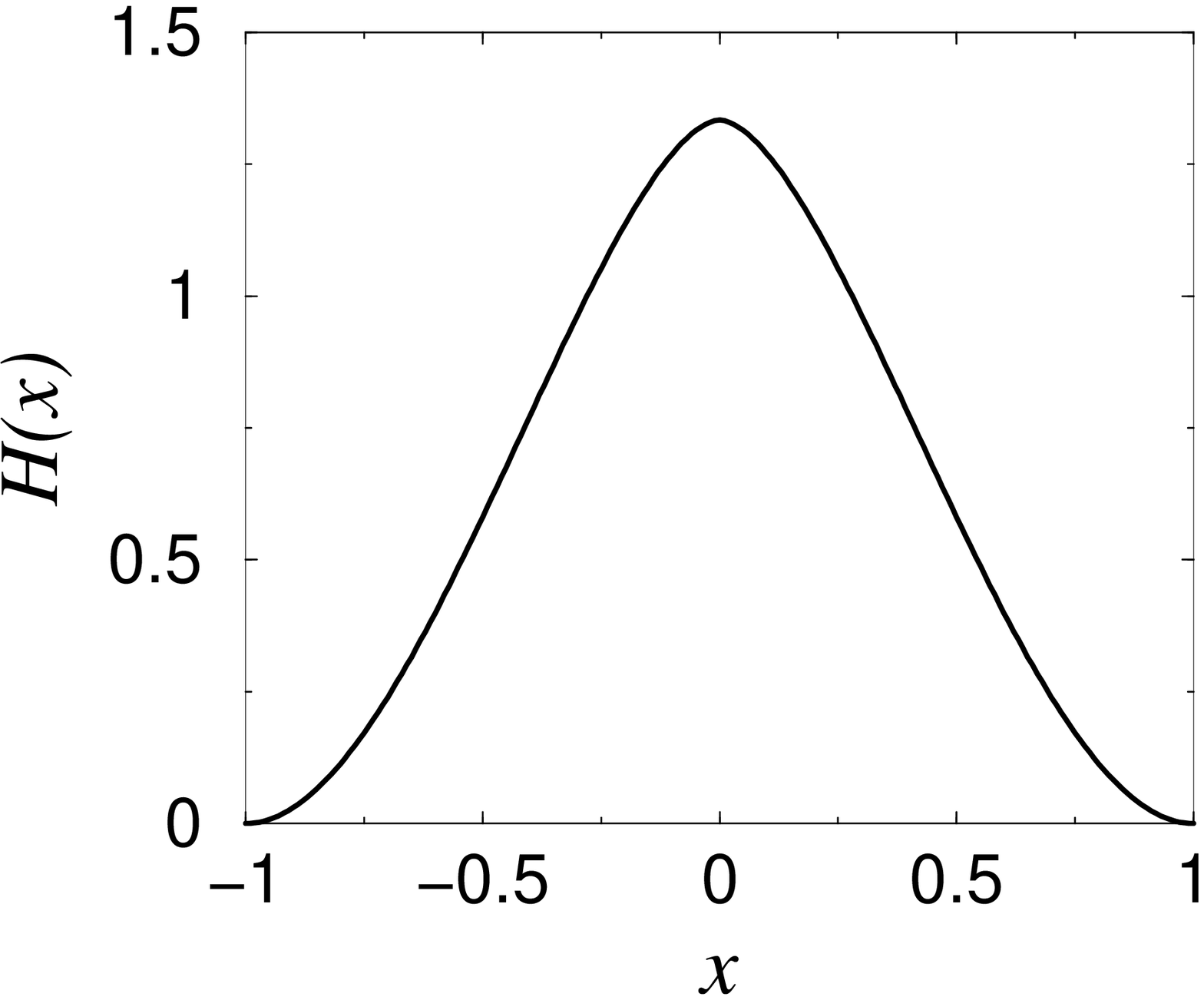}
\hskip0.3cm
\epsfxsize5.3cm\epsfbox{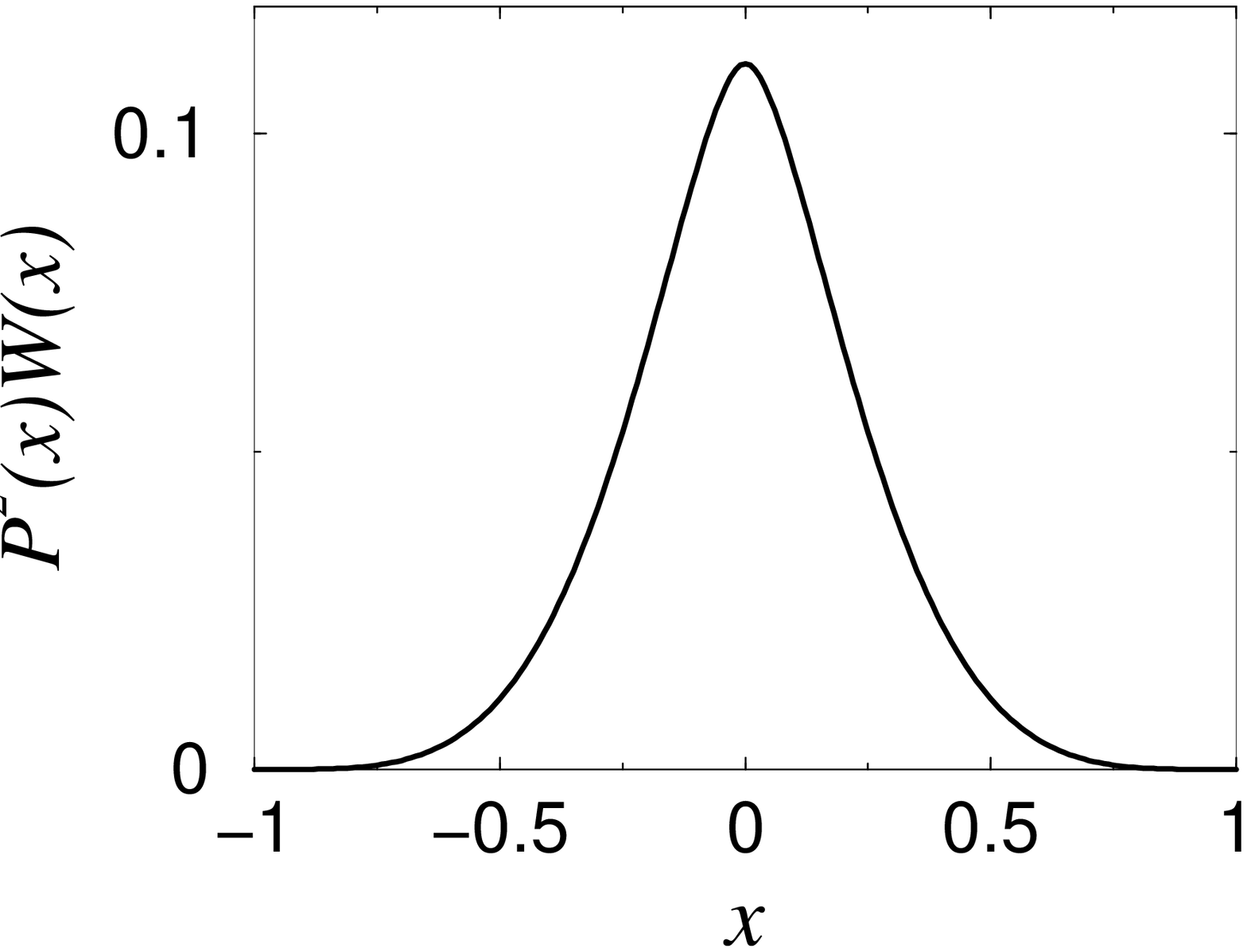}
\hskip0.5cm
\vskip .3truecm
\caption{Functions ${\cal Q}(x)$ --
Eq.~(\ref{functionQ(x)}),   and ${\cal H}(x)$ -- Eq.~(\ref{calH}), and  the product
${\cal P}^2(x){\cal W}(x)$ -- Eqs.~(\ref{calP(x)}) and (\ref{W(y)}), determining
the frequency dependence of the ``inelastic kernel''
$I(\omega)$ -- Eq.~(\ref{Iw-highT}).
 }
\label{fig-functions}
\end{figure}

In the case of higher temperatures, $\Delta\ll T \ll \omega_c,$
the main difference is related to the fact that the contribution
of a single LL to the polarization operator
is thermally smeared, yielding an extra factor $\sim \Delta/T$ as compared to
the second term of Eq.~(\ref{ballistic-realpolarization}),
as follows from Eq.~(\ref{ballistic-realpolarization-higherT}).
This changes the upper (lower) limit of integration in
$I_{\rm II}$ ($I_{\rm III}$) where $\Delta$ should be replaced by
$T$. Furthermore, in $I_{\rm II}$ one should replace $\omega_c/\Delta$
by $\omega_c/T$ in the factor related to the screening, which is equivalent
to multiplying $A$ by $\Delta/T$ in
Eqs.~(\ref{I_II-1-simple}) and (\ref{I_II-2-simple}).
This yields
\begin{eqnarray}
I_{\rm II}&\sim& {1\over a^2 R_c^2} \left({T\over \omega_c}\right)^2,
\label{I-II-result-higherT} \\
I_{\rm III}&=& {1\over a^2 R_c^2} \left({\Delta\over \omega_c}\right)^2
\ln\left({R_c T \over a \omega_c}\right).
\label{I-III-result-higherT}
\end{eqnarray}
We see that for $\Delta<T<\omega_c$ the contribution
of the ${\bf\Gamma}^{(1/qR_c)}$-term to the momentum integral
increases faster than that of ${\bf\Gamma}^{(\Delta/\omega_c)}$-term.
To evaluate this contribution more accurately, we consider the
corresponding momentum integral in the whole range of $q$ and include the
imaginary part of $\Pi({\bf q},\omega)$ into the screening (for simplicity we consider
identical layers),
\begin{eqnarray}
I(\omega)&=&\left\{
\frac{16 \Delta^2}{T^2}
{\cal P}\left({\omega \over 2\Delta}\right)
\right\}^2
\sinh^2\left({E_N-\mu \over 2T}\right)\cosh^{-6}\left({E_N-\mu \over 2T}\right)
I^{(1/qR_c)}, \label{Iw-highT}
\\
I^{(1/qR_c)} &\simeq& \int_{1/R_c}^{\infty}
{dq \over 2\pi}\left({q\over {\rm sinh} qa}\right)^2
\frac{q J_1^2(qR_c)J_0^6(qR_c)}{\{[1+A J_0^2(qR_c) ]^2 + [B J_0^2(qR_c) ]^2\}^2},
\label{I_1/qR_c}
\end{eqnarray}
where ${\cal P}(x)$ is defined in Eq.~(\ref{calP(x)}) and
\begin{eqnarray}
A&=&{2 \omega_c\over \pi T}
{\cal Q}(\omega/2\Delta){\rm cosh}^{-2}\left({E_N-\mu\over 2 T}\right),
\label{A-RePi}
\\
B&=&{2\omega_c\over \pi T} {\omega\over 2\Delta}
{\cal H}(\omega/2\Delta){\rm cosh}^{-2}\left({E_N-\mu\over 2 T}\right),
\label{B-ImPi}
\end{eqnarray}
according to Eqs.~(\ref{ballistic-realpolarization-higherT}) and
(\ref{Impi-ballistic-q-T<omegac}).
The functions ${\cal Q}(x),$
and ${\cal H}(x)$ are presented in Fig.~\ref{fig-functions}.

From the above estimates we know that the momentum integral is determined
by $q\sim \omega_c/T R_c \gg 1/R_c.$
This holds provided $A,B\sim \omega_c/T \gg 1,$ i.e.
for ${\rm cosh}([E_N-\mu]/2 T)\ll (\omega_c/T)^{1/2}.$
On the other hand, $\omega_c/T R_c\ll 1/a$ in the ballistic regime.
In this case, we can set $q^2/{\rm sinh}^2 qa = 1/a^2$ in Eq.~(\ref{I_1/qR_c}) and
set the lower integration limit to $q=0$.
Separating the fast and slow variables in Eq.~(\ref{I_1/qR_c}), we get
[$J_1(z_n)=0,\ z_n\simeq \pi n +\pi/4$]
\begin{eqnarray}
I^{(1/qR_c)} &\simeq& {1 \over 2\pi a^2 R_c^2} \sum_{n=0} z_n
\left({2\over \pi z_n}\right)^4
\int_0^\pi d\phi
\frac{\sin^2\phi\cos^6\phi}{\{[1+(2 A/\pi z_n) \cos^2\phi ]^2 + [(2 B/\pi z_n) \cos^2\phi ]^2\}^2}
\nonumber \\
&\simeq& {2 \over \pi^3 a^2 R_c^2}
\int_0^\pi d\phi \sin^2\phi\cos^2\phi
\int_0^\infty dz \frac{z}{\{[z+A]^2 + B^2\}^2}
\nonumber \\
&=&{1 \over 8\pi^2 a^2 R_c^2 B^2}
\left\{ 1-{A\over |B|}\left[{\pi\over 2} - \arctan{A\over |B|}\right]\right\}
\nonumber \\
&=&{1\over 32 a^2 R_c^2}
\left({T \over \omega_c}\right)^2
{\rm cosh}^{4}\left({E_N-\mu\over 2 T}\right) {\cal W}(\omega/2\Delta),\\
{\cal W}(x)&=&
{1 \over x^2{\cal H}^2(x)}
\left\{ 1-{{\cal Q}(x)\over |x| {\cal H}(x)}
\left[{\pi\over 2} -
\arctan{ {{\cal Q}(x)\over |x|{\cal H}(x)} }
\right]
\right\}
\label{W(y)}
\end{eqnarray}
Substituting this result into (\ref{Iw-highT}) and integrating the obtained $I(\omega)$
over frequency according to (\ref{rho-w-integral}), we arrive at
Eq.~(\ref{rho-more-leading-high-T}) of the main text.


\begin{thebibliography} {50}
\bibitem[$\dagger$]{byline}
Also at A.F.~Ioffe Physico-Technical
Institute, 194021 St.~Petersburg, Russia.

\bibitem[$\ddagger$]
{} Also at Petersburg Nuclear Physics
Institute, 188350 St.~Petersburg, Russia.
\bibitem{Eisenstein} T.J.\ Gramila, J.P.\ Eisenstein, A.H.\ MacDonald, L.N.\
Pfeiffer, and K.W.\ West, Phys.\ Rev.\ Lett.\ {\bf 66}, 1216 (1991);
Phys. Rev. B {\bf 47}, 12957 (1993).

\bibitem{Sivan} U.\ Sivan, P.M.\ Solomon, and H.\ Shtrikman, Phys.\
  Rev.\ Lett.\
{\bf 68}, 1196 (1992).

\bibitem{Hill} N.P.R.~Hill, J.T.~Nicholls, E.H.~Linfield, M.~Pepper,
  D.A.~Ritchie, A.R.~Hamilton, and G.A.C.~Jones, J. Phys.:
  Condens. Matter {\bf 8}, L557 (1996).

\bibitem{Rubel} H.~Rubel, A.~Fisher, W.~Dietsche, K.~von~Klitzing, and
  K.~Eberl, Phys.\ Rev.\ Lett.\ {\bf 78}, 1763 (1997).

\bibitem{Gramila} X.G.\ Feng, S.\ Zelakiewicz, H.\ Noh, T.J.\ Ragucci, and
T.J.\ Gramila, Phys.\ Rev.\ Lett.\ {\bf 81}, 3219 (1998).

\bibitem{Lilly98} M.P.\ Lilly, J.P.\ Eisenstein, L.N.\ Pfeiffer,
and K.W.\ West, Phys.\ Rev.\ Lett. {\bf 80}, 1714 (1998).

\bibitem{Lok} J.G.S.\ Lok, S.\ Kraus, M.\ Pohlt, W.\ Dietsche, K.\ von
  Klitzing,
W.\ Wegscheider, and M.\ Bichler, Phys.\ Rev.\ B {\bf 63}, 041305 (2001).

\bibitem{kellogg02} M.~Kellogg, I.B.~Spielman, J.P.~Eisenstein,
L.N.~Pfeiffer, and K.W.~West, Phys.\ Rev.\ Lett. {\bf 88},
126804 (2002).

\bibitem{kellogg03} M.~Kellogg, J.P.~Eisenstein, L.N.~Pfeiffer, and
  K.W.~West, Phys.\ Rev.\ Lett.\ {\bf 90}, 246801 (2003).

\bibitem{Muraki} K.\ Muraki, J.G.S.\ Lok, S.\ Kraus, W.\ Dietsche, K.\
  von Klitzing,
D.\ Schuh, M.\ Bichler, and W.\ Wegscheider, cond-mat/0311151 (2003).



\bibitem{MacDonald} L.Zheng, and A.H.\ MacDonald, Phys.\ Rev.\ B {\bf 48},
8203 (1993).

\bibitem{Bonsager} M.C.\ B{\o}nsager, K.\ Flensberg, B.Y.\ Hu, and
  A.-P.\ Jauho, Phys.\ Rev.\ Lett. {\bf 77}, 1366 (1996);
Phys.\ Rev.\ B {\bf 56}, 10314 (1997).

\bibitem{Khaetskii} A.V.\ Khaetskii and Yu.V.\ Nazarov, Phys.\ Rev.\ B
  {\bf 59}, 7551 (1999).


\bibitem{Oppen} F.\ von Oppen, S.H.\ Simon, and A.\ Stern, Phys.\ Rev.\ Lett.\
{\bf 87}, 106803 (2001).

\bibitem{Kamenev} A.\ Kamenev and Y.\ Oreg, Phys.\ Rev.\ B {\bf 52},
  7516 (1995).

\bibitem{Flensberg95} K.\ Flensberg, B.Y.\ Hu, A.-P.\ Jauho, and J.M.\
Kinaret, Phys.\ Rev.\ B {\bf 52}, 14761 (1995).

\bibitem{Ando} T.\ Ando and Y.\ Uemura, J.\ Phys.\ Soc.\ Jpn.\ {\bf
    36}, 959 (1974);
T.\ Ando, {\it ibid} {\bf 37}, 1233 (1974).

\bibitem{Chalker} K.A.~Benedict and J.T.~Chalker, J. Phys. C {\bf 19},
  3587 (1986). For a finite correlation length $d$ of disorder, the
  SCBA is justified under the condition $\ell \gg d$, see M.E.~Raikh
  and T.V.~Shahbazyan, Phys.\ Rev.\ B {\bf 47}, 1522 (1993);
  B.~Laikhtman and E.L.~Altshuler, Ann.\ Phys.\ {\bf 232}, 332 (1994).

\bibitem{unpublished} I.V.\ Gornyi, A.D\ Mirlin, and F.\ von Oppen,
  unpublished.

\bibitem{Aleiner-drag} B.N.\ Narozhny, I.L.\ Aleiner, and A.\ Stern,
Phys.\ Rev.\ Lett. {\bf 86}, 3610 (2001).

\bibitem{bonsager-comment} We remark that it was erroneously claimed in Ref.\
\onlinecite{Bonsager} that the vertex corrections are negligible for $n\neq n'$.

\bibitem{foot-vert} Keeping the LL indices in scalar vertex corrections,
it can be checked that the factor
${\rm Re}\, [ \gamma^{-+}({\bf q},\omega)-\gamma^{++}({\bf q},\omega)]$
in Eq.~(\ref{leading-gamma-finiteTw})
involves only the vertex corrections of type $\gamma^{\mu\nu}_{N,N\pm1}$.
Therefore, neglecting such vertex corrections~\cite{bonsager-comment}
one loses the leading term (\ref{leading-gamma-finiteTw}).

\bibitem{Aleiner} I.L.\ Aleiner and L.\ Glazman, Phys.\ Rev.\ B {\bf
    52}, 11296 (1995).

\bibitem{Vavilov} M.G.\ Vavilov and I.L.\ Aleiner,
Phys.\ Rev.\ B {\bf 69}, 035303 (2004).

\bibitem{foot-CF} Though the theoretically predicted~\cite{CF-Stern,CF-Sakhi,CF-KM}
anomalous $T^{4/3}$ dependence of $\rho^D_{xx}$
at half-filling of the lowest LL
is reasonably well confirmed by the experimental data~\cite{Lilly98,Muraki} for the
weak-coupling regime, the observed~\cite{Lilly98}
dependence of the drag on the magnetic field
around the half-filling seems to disagree strongly with theory.

\bibitem{CF-Stern} I.\ Ussishkin and A.\ Stern, Phys.\ Rev.\ B {\bf 56}, 4013 (1997).

\bibitem{CF-Sakhi}  S.\ Sakhi, Phys.\ Rev.\ B 56, 4098 (1997).

\bibitem{CF-KM} Y.-B.\ Kim and A.J.\ Millis, Physica E (Amsterdam) {\bf 4}, 171 (1999).

\bibitem{Shimshoni} The Coulomb drag at the quantum Hall transition
  was considered in
E.~Shimshoni and S.L.~Sondhi, Phys.\ Rev.\ B {\bf 49},
  11484 (1994). However, this work suffers from the same deficiency as
  Refs.~\onlinecite{Bonsager,Khaetskii}: its starting point is a formula for
  drag which misses the
  contributions related to particle-hole asymmetry induced by magnetic
  field.

\end{thebibliography}
\end{document}